\documentclass[12pt,preprint]{aastex}
\usepackage{epsfig}
\usepackage{lscape}

\makeatletter

\newcommand{\Rmnum}[1]{\expandafter\@slowromancap\romannumeral #1@}
\makeatother

\def\Lsun{\hbox{\it L$_\odot$}}

\def\Msun{\hbox{\it M$_\odot$}}

\def\sgrastar{\mbox{Sgr A$^*$}}
\def\cxo{\textit{Chandra X-ray Observatory$~$}}
\newcommand{\etal}{\mbox{et al.}}

\newcommand{\program}[1]{{\tt {#1}}}
\def\program{\texttt}

\begin{document}
\title{Near-Infrared Counterparts to \textit{Chandra} X-ray Sources Toward the Galactic Center. {\sc II}. Discovery of Wolf-Rayet Stars and O Supergiants}

\shorttitle{}

\author{J.~ C. Mauerhan\altaffilmark{1}, M.~ P. Muno\altaffilmark{2}, M.~R. Morris\altaffilmark{3},  S.~R. Stolovy\altaffilmark{4},  A. Cotera\altaffilmark{5}}

\altaffiltext{1}{Spitzer Science Center, California Institute of Technology, Pasadena, CA 91125, USA; mauerhan@ipac.caltech.edu}
\altaffiltext{2}{Space Radiation Laboratory, California Institute of Technology, Pasadena, CA 91125, USA}
\altaffiltext{3}{Department of Physics and Astronomy, University of California, Los Angeles, CA 90095-1547, USA}
\altaffiltext{4}{Spitzer Science Center, California Institute of Technology, Pasadena, CA 91125, USA}
\altaffiltext{5}{SETI Institute, 515 N. Whisman Rd., Mountain View, CA, USA}

\begin{abstract}
We present new identifications of  infrared counterparts to the population of hard X-ray sources near the Galactic center detected by the \textit{Chandra X-ray Observatory}. We have spectroscopically confirmed 16 new massive stellar counterparts to the X-ray population, including nitrogen-type (WN) and carbon-type (WC) Wolf-Rayet stars, and O supergiants.  These discoveries increase the total sample of massive stellar X-ray sources in the Galactic center region to 30 (possibly 31). For the majority of these sources, the X-ray photometry is consistent with thermal emission from plasma having temperatures in the range of $kT=1$--8 keV or non-thermal emission having power-law indices in the range of $-1\lesssim\Gamma\lesssim3$, and X-ray luminosities in the range of $L_{\textrm{\scriptsize{X}}}\sim10^{32}$-- $10^{34}$ erg s$^{-1}$ (0.5--8.0 keV). Several sources have exhibited X-ray variability of several factors between observations. These X-ray properties are not a ubiquitous feature of single massive stars but are typical of massive binaries, in which the high-energy emission is generated by the collision of supersonic winds, or by accretion onto a compact companion. However, without direct evidence for companions, the possibility of intrinsic hard X-ray generation from single stars cannot be completely ruled out.   The spectral energy distributions of these sources exhibit significant infrared excess, attributable to free-free emission from ionized stellar winds, supplemented by hot dust emission in the case of the WC stars. With the exception of one object located near the outer regions of the Quintuplet cluster, most of the new stars appear isolated or in loose associations. Seven hydrogen-rich WN and O stars are concentrated near the Sagittarius B H {\sc ii} region, while other similar stars and more highly evolved hydrogen-poor WN and WC stars lie scattered within $\approx$ 50 pc, in projection, of Sagitarrius A West. We discuss various mechanisms capable of generating the observed X-rays and the implications these stars have for massive star formation in the Galaxy's Central Molecular Zone.
\end{abstract}

\section{Introduction}
The Galactic center harbors the largest concentration of X-ray sources in the sky. Indeed, a total of 9017 X-ray sources have been detected in deep observations of the central $2^{\circ}\times0\fdg8$ of the Galaxy with the {\cxo} (Muno et al. 2009). The majority of these sources are highly absorbed by interstellar gas and dust, indicative of distances near the Galactic center ($D\approx8$ kpc; Reid 1993).  A wide variety of stellar phenomena comprise the Galactic center X-ray population, including accreting magnetic white dwarfs in cataclysmic variables (CVs; e.~g., see Muno et al. 2003) and symbiotic binaries; accreting neutron stars and black holes in low- and high-mass X-ray binaries (e.~g., see Pfahl, Rappaport \& Podsiadlowski 2002); Wolf-Rayet and O-supergiant binaries with colliding supersonic winds; and, perhaps, rogue compact objects accreting from the dense Galactic center medium (Agol \& Kamionkowski 2002). All such objects are rare among stellar populations, and involve exotic forms of matter and high-energy radiation mechanisms, of which our knowledge is fragmentary.  Thus, the characterization of the Galactic center X-ray population is an important step toward improving our understanding of the end stages of stellar evolution, and the `zoology' of X-ray sources contained within the only galactic nucleus we are currently able to resolve in such detail.

X-ray properties alone, however, are insufficient to characterize the Galactic center X-ray population. The detection of counterparts at other wavelengths is necessary, although such a pursuit is limited by the high extinction suffered by starlight traversing 8 kpc from the Galactic center to Earth ($A_V\approx30$ mag). Fortunately, the brightness distribution of infrared counterparts can be used to constrain the relative contributions of low- and high-mass objects to the sample. For instance, CVs with cool dwarf donors in the Galactic center should have very faint IR counterparts with $K\gtrsim22$ mag, while main-sequence stars earlier than B0{\sc V}, and Wolf-Rayet stars, will have $K\lesssim15$ and $\lesssim12$ mag, respectively. It was demonstrated in Mauerhan et al. (2009; hereafter referred to as Paper I) that only $\approx6$\%$\pm2$\% of the 6760 absorbed X-ray sources have real infrared counterparts with $K_s\le15.6$ mag. Although this result is consistent with earlier studies that concluded the population to be widely dominated by CVs (Muno et al. 2003, Laycock et al. 2005), it suggests, nonetheless, that $\approx100$--300 of the X-ray sources in our sample have real infrared counterparts within a brightness range consistent with that of the population of late-type giants and hot massive stars observable in the Galactic center. Indeed, several X-ray sources with hot supergiant counterparts have already been discovered, including O supergiants, and nitrogen-type (WN) and carbon-type (WC) Wolf-Rayet stars (Muno et al. 2006, Mikles et al. 2006, Mauerhan, Muno \& Morris 2007, Hyodo et al. 2008).

The successful detection of massive-star counterparts to \textit{Chandra} sources is not surprising, given the environmental conditions at the Galactic center. The region boasts the highest star formation rate density in the Milky Way (10$^{-7}$ \Msun~yr$^{-1}$ pc$^{-3}$; Figer et al. 2004), owing to the immense reservoir of molecular gas that occupies the Central Molecular Zone (e.g., see Morris \& Serabyn 1996). This material is vigorously forming massive stars near the Sagitarrius B H {\sc ii} region (e.~g., see dePree et al. 1998), and within the last several Myr has produced at least three extraordinary stellar clusters that are among the most massive and dense in the Galaxy: the Arches and Quintuplet (Nagata et al. 1995; Cotera et al. 1996, Figer et al. 1999, 2002), and the Central parsec cluster (Krabbe et al. 1995).  These clusters are rich in Wolf-Rayet stars and O supergiants. Such stars are typically sources of soft, thermal X-rays with $kT\lesssim1$ keV, while only a small subset of massive stars (those in close binaries) will produce hard X-rays with $kT>1$--2 keV. Since the softer X-ray photons are heavily absorbed by interstellar gas and dust, only the harder sources are likely to be detected at the Galactic center. This explains why the Arches and Quintuplet clusters contain only several detectable X-ray sources among their hundred or so massive stars (Yusef-Zadeh et al. 2002; Law \& Yusef-Zadeh 2004; Wang et al. 2006). Outside of these clusters, the relatively isolated X-ray-emitting supergiants discovered by Mauerhan, Muno \& Morris (2007), which reside within $\approx10$ projected pc of the Arches and Quintuplet clusters, may have been dynamically ejected from one of these systems. Alternatively, these and other isolated massive stars throughout the Galactic center may be the products of an isolated mode of massive star formation operating in the region, in tandem with the formation of stellar clusters. Thus, the identification of infrared counterparts to the \textit{Chandra} X-ray population can provide insight into the evolution of stellar clusters, and highlight previously unknown regions of massive star formation in the Galactic center. 

In this work, we present the results of our pursuit of counterparts via infrared spectroscopy. In Section 2, we list the various facilities and instrumentation used for this work and describe our observations.  In Section 3, the spectra of the newly found massive stars are presented, including new spectra of several previously discovered objects. In Section 4, the stellar parameters of confirmed sources are examined via their photometry and infrared spectral energy distributions (SEDs). In Section 5, the X-ray photometric properties of the massive stars are listed and compared with the greater field population of X-ray sources in the Galactic center. In Section 6, we discuss potential hard-X-ray emission mechanisms, consider the spatial distribution, origin, and formation mode of these stars, and discuss the implications for the greater population of massive stars in the Central Molecular Zone.

\section{Observations}
We performed infrared spectroscopy of candidate counterparts to X-ray sources that were selected via the cross-correlation of the \textit{Chandra} catalog of Muno et al. (2009) and the SIRIUS and Two Micron All Sky Survey (2MASS; Cutri et al. 2003) near-infrared catalogs. The details of the sample and the cross-correlation analysis are presented in Paper I. We targeted sources that (1) have highly absorbed soft X-ray fluxes and reddened near-infrared photometry (the so-called \textit{red} infrared matches to \textit{hard} X-ray sources from Paper 1), which is the characteristic of objects lying at the distance to the Galactic center, (2) lie at angular distances $>$7\arcmin~from \sgrastar, thus avoiding the high stellar confusion and large number of accidental infrared/X-ray matches that lie within this radius (3) have $K_{s}<12$ mag, and (4) have excellent X-ray astrometry, with positional uncertainties $\le$1\arcsec.  We have currently obtained spectra of 52 infrared matches to X-ray sources. We observed in the $K$-band since it contains many diagnostic features of both hot and cool stars (e.g., see Hanson et al. 1996, 2005; Morris et al. 1996; Figer, McLean \& Najarro 1997), and suffers the least amount of extinction in the near infrared. Eighteen sources, including those from Mauerhan, Muno \& Morris (2007), exhibit characteristics of massive stars (see Section 3), while the remainder exhibit CO band-head absorption features near $\lambda2.3$ ${\micron}$, and sodium absorption features near $\lambda2.21$ ${\micron}$, typical of cool, late-type stars (spectra not presented here). In general, we assumed that a late-type match to an X-ray source was a spatial coincidence, since late-type stars vastly outnumber hot stars in the field. Thus, late-type counterparts will not be presented in this paper. However, we note that symbiotic X-ray binaries consisting of late-type giants or supergiants with accreting compact companions could be indistinguishable from random coincidences if emission-line accretion signatures were too weak to be detected in our spectra. Thus, we cannot completely rule out the presence of symbiotic stars in the late-type sample. 

We do not claim that the true statistics of infrared counterparts to X-ray sources are indicated by the number of massive-star detections presented below; there are several reasons for this: our spectroscopy campaign began before the completion of the \textit{Chandra} survey and the final source catalog, which in the end allowed us to refine the X-ray point-source positions and uncertainties, and eliminate spurious detections (the method of refining the X-ray astrometry is discussed in Muno et al. 2009). Thus, several matches for which we had obtained prior spectroscopy did not remain in the final candidate counterpart list that was based on the refined X-ray catalog. In fact, only 15 of the 35 sources which we confirmed spectroscopically to be late-type stars remained in the final list. Furthermore, candidate counterparts were also given priority for spectroscopy based upon their location in particular regions of interest. For instance, if a candidate's position on the sky places it near an interesting diffuse structure, such as a bubble-like or shell-like feature in the \textit{Spitzer}/IRAC $\lambda$8 \micron~images of the region, then it was given priority over other candidate counterparts in the list. The motivation for this additional selection criterion lies in the fact that intense winds from massive stellar X-ray sources can sweep up and heat interstellar material.  Thus, we interpreted the spatial association of candidate infrared/X-ray counterparts with diffuse mid-infrared structures as indicative of a higher likelihood that the X-ray source is associated with a massive star. Although this technique probably increased our success rate, it almost certainly imposed a selection effect that hampers our ability to statistically analyze our spectroscopic success rate for finding counterparts to the \textit{Chandra} population. 

Table 1 lists the X-ray sources and associated massive-star counterparts that we targeted for spectroscopy, along with the instrument used, and the dates of the observations. The table also includes information on seeing and sky conditions. All sources are presented with their full \textit{Chandra} designation in Table 1, and are hereafter referred to in the text by an abbreviated version (e.g., CXOGC J174532.7$-$285617 will hereafter be referred to as X174532.7).  

\subsection{NASA Infrared Telescope Facility (IRTF) and SpeX}
Spectra of four stars were obtained using the SpeX medium-resolution spectrograph on the 3 m Infrared Telescope Facility (IRTF) telescope (Rayner et al. 2003), located on the summit of Mauna Kea in Hawaii. For X174656.3 and X174711.4, SpeX was used in the high-throughput, low-resolution prism mode, which provides coverage of the entire \textit{JHK} bandpass, although we will only include the $K$-band in this work. A slit width  of 0\farcs3 was used, providing a spectral resolution of $R\approx250$. For the infrared counterparts to X174645.2 and X174550.6, medium-resolution spectra were obtained using the short cross-dispersed mode (SXD) of SpeX, with slit widths of 0\farcs5 and 0\farcs3 for the respective stars, providing spectral resolutions of $R\approx1200$ and $R\approx2000$ in the $K$-band. All spectra were acquired in an ABBA nodding sequence in order to subtract the sky background, and to suppress the contribution of bad pixels. The spectra were reduced and extracted using the \program{IDL}-based software package \program{Spextool}, specially designed for the reduction of data obtained with SpeX on the IRTF (Cushing et al. 2004).  For telluric correction, spectra of the A0V standard star HD162220 were obtained and applied to all of the low-resolution data, while the A0{\sc V} star HD155379 was used for the medium-resolution spectra. Telluric correction was executed using the IDL package \program{xtellcor} (Vacca et al. 2003), which applies and removes model H {\sc i} absorption lines from the A0{\sc V} standard star before application to the science data.

\subsection{The United Kingdom Infrared Telescope (UKIRT)}
The 3 m United Kingdom Infrared Telescope (UKIRT) on Mauna Kea was used to obtain spectra of eight infrared counterparts: X174519.1, X174532.7, X174537.3, X174628.2, X174703.1, X174712.2, X174713.0, and X174725.3. The UKIRT 1--5 {\micron} Imager Spectrometer (UIST; Ramsay et al. 2004) was used in service mode as part of the UKIRT Service Programme. The short-K grism and 4-pixel slit (0\farcs48) were used, which provided a spectral resolution of $R\approx2133$ and a wavelength range of 2.01--2.26 $\micron$. Spectra were acquired in an ABBA nodding sequence for sky subtraction. Basic reductions were executed by the \program{Starlink ORACDR} pipeline and the spectra were extracted using the \program{IRAF} routine \program{APALL}. The telluric spectrum of HD162220 was applied to all UKIRT/UIST data using the \program{xtellcor\_general} program, a generalized version of the same program used for IRTF SpeX data (used for all stars not observed at the IRTF).

\subsection{The Anglo-Australian Telescope and IRIS2}
Spectra of X174516.1, X174616.6,  and X174617.7  were obtained using the 4.1 m Anglo-Australian Telescope (AAT) on Siding Spring Mountain (Mount Woorat) in New South Wales, Australia. The IRIS2 instrument (Tinney et al. 2004) provided a spectral resolution of $R\approx2400$ in the $K$-band, using the 1\arcsec~slit. The data images were reduced using the \program{Starlink ORACDR} pipeline. The spectra were extracted using the \program{IRAF} routine \program{APALL}. The telluric spectrum was obtained from observations of the F0V star BS6441. 

\subsection{The Southern Observatory for Astrophysical Research and OSIRIS}
Spectra of X174502.8, X174508.9 and X174516.7 were obtained at the Southern Observatory for Astrophysical Research (SOAR), located on Cerro Pachon in Chile. The Ohio State Infrared Imager Spectrometer (OSIRIS; Depoy et al. 1993) was used in high-resolution longslit mode, which provided $R$$\approx$3000  in the $K$-band. Stellar spectra were acquired in a slit-scan sequence of five positions separated by 5\arcsec~each. The five spectral images were median combined with a suitable bad-pixel rejection algorithm to produce a master sky spectrum, which was subtracted from each individual exposure. The A0V standard HD155379 was used for telluric calibration. 

\section{Spectral Classification}
In the following subsections, we estimate the spectral types of the infrared counterparts to Galactic center X-ray sources.  All of the spectral criteria we use to classify these stars were adopted from previous spectroscopic studies of massive stars in the near-infrared, conducted by Morris et al. (1996), Hanson et al. (1996, 2005), Figer, McLean \& Najarro (1997), and  Martins et al. (2008). The central wavelengths of all spectral lines referred to in the text are adopted from Figer, McLean \& Najarro (1997, their Table 2). 

\subsection{Late O Supergiants}
Figure 1 shows the $K$-band spectra of the infrared counterparts to X174537.3, X174502.8, and the Quintuplet member X174616.6 (star qF344 of Figer, McLean \& Morris 1999).  Each of these stars exhibits absorption lines of Br$\gamma$ at $\lambda$2.1661 {\micron} and He {\sc i} at $\lambda$2.058, $\lambda$2.113, and $\lambda$2.1647 {\micron}, which are typical features of late O and early B supergiants. In addition, each star also exhibits a weak emission feature of N {\sc iii}, which appears at $\lambda$2.115 {\micron}, on the red side of the He {\sc i} absorption line.  The main feature distinguishing the stars of Figure 1 is the varying depth of the Br$\gamma$ line, which tends to increase with later spectral type. The Br$\gamma$ feature is weakest for X174502.8, deepest for X174537.3, and is of intermediate strength for X174616.6. The latter star exhibits a weak absorption line of He {\sc ii} as well, which indicates that it is somewhat hotter and of slightly earlier spectral type than X174537.3 and X174502.8.  Based on a comparison of these spectra with the OB stars of Hanson et al. (1996, 2005) , we classify X174502.8 and X174537.3 as supergiants within the O9I--B0I range, and X174616.6 as an O8--O9I star.

We note that the X-ray detection of X174502.8 is questionable. The source detection was flagged as spurious and not included in the catalog of Muno et al. (2009), but owing to the source's position near an interesting shell-like feature in \textit{Spitzer} $\lambda$8 \micron~images, it was targeted for spectroscopy nonetheless. This implies that either some legitimate X-ray detections near the threshold of significance have been discarded by our selection algorithm, or the spatial correlation of the OB star with a truly spurious X-ray detection was a very unusual coincidence.  

\subsection{Early O Supergiants}
The infrared counterparts of X174532.7, X174628.2, X174703.1, and X174725.3  all exhibit very similar $K$-band spectra, presented in Figure 2. The spectra are dominated by a complex of blended emission lines near $\lambda$2.112--2.115 {\micron} that is dominated by He {\sc i} and N {\sc iii}, and might also may contain contributions from C {\sc iii} and O {\sc iii}.  Blueward of this line complex is weaker emission lines of C {\sc iv} at $\lambda$2.069 and $\lambda$2.078 {\micron}. Br$\gamma$ is seen in emission in X174628.2, although it may be nebular feature, while there is marginal detection of Br$\gamma$ absorption in X174532.7. Both X174532.7 and X174703.1 exhibit He {\sc ii} absorption at $\lambda$2.189 {\micron}, which is either absent or below the noise level in the other two stars. All of these features are consistent with the $K$-band spectra of early O supergiants, specifically O4--6{\sc I} stars (e.~g., see Martins et al. 2008) and we classify them as such.

The Arches cluster of the Galactic center is rich with stars of O4--6{\sc I} spectral type (Figer et al. 2002, Najarro et al. 2004, Martins et al. 2008). For comparison, accompanying the spectra in Figure 2 is a model spectrum for the O4--6{\sc I} star F18 of the Arches cluster from Martins et al. (2008). This model was computed using the stellar atmosphere code \program{CMFGEN} (Hillier \& Miller 1998). The model star has the following values for effective temperature, luminosity, mass-loss rate, and terminal wind velocity of these stars: $T_{eff}=36900$ K, $L=10^{5.9} L_\odot$,  $\dot{\mbox{M}}=3.2\times10^{-6} {M}_{\odot}$ yr$^{-1}$, and  $v_{\infty}=2150$ km s$^{-1}$. It is reasonable to assume that the stars in Figure 2 have similar properties.
  
 \subsection{WNh and OIf$^+$ Stars}
The $K$-band spectra of X174656.3, X174711.4, and X174617.0 are presented in Figure 3. The spectra of these stars are dominated by emission lines of Br$\gamma$ and the $\lambda$2.112--2.115 {\micron} complex of He {\sc i}, N {\sc iii}, C {\sc iii} and O {\sc iii}. The spectrum of X174617.0 was first presented in Mauerhan, Muno \& Morris (2007; hereafter referred to as MMM07), where it was classified as an O6If$^+$ star, but we present a refined version of it here. The Br$\gamma$ line of this particular star exhibits an asymmetry on its blue side owing to a contribution from He {\sc i}, although the detectability of this feature may be due to the fact that the spectrum of X174617.0 has significantly higher spectral resolution than the other stars in Figure 3. An absorption line of He {\sc i} at $\lambda$2.058 {\micron} is present for all three stars, although the detection is marginal for X174711.4. Weak N {\sc iii} emission appears at $\lambda$2.247 {\micron} in the low-resolution spectra of X174711.4 and X174656.3, while this line appears as a strong doublet feature in the higher resolution spectrum of X174617.0. Weak N {\sc v} emission is also observed in all three stars, although its detection in the higher resolution spectrum of X174617.0 is the most secure.  X174617.0 also exhibits a weak He {\sc ii} line ($\lambda$2.189 {\micron}) in absorption.  

These features are consistent with O4--6{\sc I}f$^+$ and core-H burning WNL stars of the hydrogen-rich variety (WNh, Smith et al. 1996). O{\sc I}f$^+$ and WN8--9h stars with relatively weak lines can be very difficult to distinguish from one another using near-infrared spectroscopy alone, as these two spectral types occupy opposite ends of a continuous morphological sequence (Bohannan \& Crowther 1999). WN8--9h stars are usually distinguished by He {\sc ii} ($\lambda$2.189 {\micron}) in emission or as a P Cygni profile. However, several O{\sc I}f$^+$ have been shown to exhibit this line in emission as well, while several WN8--9h stars are known to exhibit this line in absorption (Conti et al. 1995; Bohannan \& Crowther 1999). An additional classification criterion, more relevant for the $K$ band, is provided by the Br$\gamma$ emission line, which is usually stronger than the $\lambda$2.112--2.115 {\micron} complex for WN8--9h stars, relatively weak in O{\sc I}f stars, and of comparable strength in O{\sc I}f$^+$ stars; based upon this criterion, we classify X174656.3 and X174711.4  as WN8--9h stars, and maintain the original O6If$^+$ classification from MMM07 for X174617.0. 

WN8--9h stars are also well represented in the Arches cluster. A  model spectrum of the WN8-9h star F7 from Martins et al. (2008) is included in Figure 3 for comparison with the other WN8--9h stars. The basic parameters of this model are $T_{eff}=32,900$ K, $L=10^{6.3}L_\odot$,  $\dot{\mbox{M}}=2.5\times10^{-5} ~M_{\odot}$ yr$^{-1}$, and  $v_{\infty}=1300$ km s$^{-1}$. 

Figure 4 presents the $K$-band spectra of X174712.2, X174713.0, and X174516.7. These stars are distinguishable from the WN8--9h stars of Figure 3 by their more prominent He {\sc ii} emission at $\lambda$2.189 {\micron}, indicative of slightly earlier spectral types in the range of WN7--8h (e.~g., see Martins et al. 2008). The higher signal-to-noise spectrum of X174712.2 also exhibits emission features of N {\sc v} near $\lambda$2.10 {\micron}, and C {\sc iv} emission near $\lambda$2.078 {\micron}. In conclusion, we classify  X174712.2, X174713.0, and X174516.7 as WN7--8h stars.

\subsection{Hydrogen-Deficient WN Stars}
Figure 5 presents $K$-band spectra of WN stars X174550.6, X174508.9, X174555.3, and X174522.6. These stars have hydrogen-deficient spectra containing very broad, flat-topped emission lines of He {\sc i} and He {\sc ii}, indicative of fast, extended stellar winds. The discovery of X174550.6 was first reported in Cotera et al. (1999), and it was first identified as an X-ray source in Muno et al. (2006). We present a new, higher resolution $K$-band spectrum of this star and examine its spectral type. 

For X174550.6 we measure EW($\lambda$2.189 {\micron})/EW($\lambda$2.1661 {\micron})=0.52 and EW($\lambda$2.189 {\micron})/EW($\lambda$2.112 {\micron})=0.55, which is consistent with stars of subtype WN7, according to Figer, McLean \& Najarro (1997); this classification updates the earlier WN6 classification in Cotera et al. (1999). By direct comparison, the spectrum of X174508.9 is very similar to that of X174550.6, also exhibiting EW ratios consistent with WN7 classification. Slight differences in the spectrum of X174508.9 include a stronger He {\sc ii} emission at $\lambda$2.189 and $\lambda$2.314 {\micron}, and a deeper P Cygni absorption component from He {\sc i} at $\lambda$2.058 {\micron}.  The He {\sc i} absorption component for X174508.9 also has larger blueshift than that of X174550.6, implying that the wind of X174508.9 may be faster and more extended than that of X174550.6. 

The counterpart to X174555.3 was first presented in MMM07, where it was classified as a WN6b star. However, a refined spectral analysis of this star demonstrates that the original telluric correction and continuum fitting resulted in an underestimate of the He {\sc i} ($\lambda$2.0587 {\micron}) emission-line flux. The more reliable result presented here exhibits a stronger emission line at this wavelength, with a blueshifted absorption profile from the same He {\sc i} transition superimposed on the emission. This absorption line forms in the outer parts of the stellar wind where He {\sc ii} has recombined, and so provides an estimate of the wind velocity there. We measure the absorption line centroid at {$\lambda$}=2.0425 {\micron}, which implies a wind velocity of $2400\pm100$ km s$^{-1}$, typical for WN5--6 stars (Crowther \& Smith 1996). Furthermore, the broad emission feature near $\lambda$2.11 {\micron} appears to extend further into the blue end of the line, when compared with the same feature in the WN7 stars X174550.6 and X174508.9. We attribute this to N {\sc v} emission, which is a characteristic of WN4--5 stars (Crowther et al. 2006a). Therefore we relax our previous classification of WN6b to WN5--6b for X174555.3. Finally, we classify X174522.6 as a WN5--6b star as well, owing to the strong similarity between its spectrum and that of X174555.3.

\subsection{Weak-lined WC9 Stars}
Figure 6 presents the $K$-band spectra of X174519.1, X174617.1 and X174645.2. The former two stars exhibit broad emission lines from the complex of He {\sc i}, N {\sc iii}, C {\sc iii} and O {\sc iii} ($\lambda$2.112--2.115 {\micron}), He {\sc i} and He {\sc ii} ($\lambda$2.058 \micron), C {\sc iii} and C {\sc iv} (near $\lambda$2.07--2.08 {\micron}), and weak He {\sc ii} emission near $\lambda$2.189 {\micron}. These spectral characteristics are consistent with those of late carbon-type (WC) Wolf-Rayet stars, specifically WC9 stars, whose values of EW($\lambda$2.08 {\micron})/EW($\lambda$2.112 {\micron}) are below 1 (e.~g., see the spectrum of WR 112 in Figer, McLean \& Najarro 1997). However, the peaks of the strongest emission lines are only $\approx10\%$ of the continuum flux. WC9 stars typically exhibit much stronger lines. However, WC stars associated with hot thermal dust emission (WC9d stars) typically exhibit relatively weak emission lines in the $K$-band, much like those of X174519.1 and X174617.7 (e.~g., see Crowther et al. 2006a, their Figure 11). This is presumed to be the result of a bright thermal dust continuum which competes with the line emission of stellar wind.

The lower spectrum in Figure 6 is for X174645.2, and appears completely featureless within the noise level. Based on this star's infrared SED and hard X-ray emission, it was identified by Hyodo et al. (2008) a dust-enshrouded, late-type WC binary, also known as a DWCL star (Williams et al. 1987), the same spectral type as the Quintuplet proper members (e.~g., see Figer, McLean \& Morris 1999). Figure 6 provides the first near-infrared spectrum of this star, and it is consistent with the DWCL spectral type based on its featureless character. Although the weakness or absence of emission lines in the near-infrared spectra of late-type WC stars is mainly attributable to dilution by continuum emission from hot dust, line dilution may also result from the stellar continuum emission from a very luminous companion. Later, in Section 4, we will present additional evidence for the presence of hot dust associated with X174519.1 and X174617.7.

 \subsection{CXOGC J174516.1$-$290315}
X174516.1 is one of the brightest emission-line stars known in the Galactic center ($K_{s}=7.84$ mag). This source was first identified as a young stellar object by Felli et al. (2002), and it was subsequently characterized as an Ofpe or LBV emission-line star by Muno et al. (2006), who also showed this star to have a 2.3 mJy radio counterpart at $\nu$8.4 GHz. Stars of these spectral types are hydrogen rich, yet represent the beginning of post main-sequence evolution for the most massive stars (Morris et al. 1996).  Figure 7 shows our $K$-band spectrum of X174516.1 at medium resolution. Strong emission lines of Br$\gamma$ and He {\sc i} are accompanied by weak low-ionization emission lines of Mg {\sc ii}. The $\lambda$2.058 {\micron} line is the strongest emission feature and appears as a P-Cygni profile, while the He {\sc i} line at $\lambda$2.112 {\micron}, also exhibiting a P-Cygni profile, is much weaker and may be blended with weak emission from N {\sc iii}.  There also appears to be a weak P-Cygni feature of He {\sc ii} near $\lambda$2.19 {\micron}.  Based on the above characteristics, X174516.1 appears to be a member of Ofpe/WN9 class (e.~g., see Morris et al. 1996), also resembling several stars that lie near the central parsec of the Galactic center, namely IRS16C, IRS16NW, and IRS33E, the spectra of which were modeled in Martins et al. (2007). Owing to the strong similarity of our spectra to these stars, we classify X174516.1 as an Ofpe/WN9 star.

From the P-Cygni profile of He {\sc i} $\lambda$2.058 {\micron}, we may infer the velocity at the radius in the wind where He {\sc i} becomes an absorption line. The trough in the $\lambda$2.0587 {\micron} feature is centered at 2.0519 {\micron} and the emission peak is centered at $\lambda$2.0541 {\micron}, so the implied velocity in the absorption zone is $\approx320$ km s$^{-1}$.

\section{Stellar Properties}
\subsection{Bolometric Luminosity}
The near-infrared photometry of each massive star is presented in Table 2, which also includes mid-infrared photometry from the \textit{Spitzer} Infrared Array Camera (IRAC) point-source catalog of the Galactic Center (Ram{\'{\i}}rez et al. 2008). To derive the absolute photometry, we must first determine the effects of interstellar extinction. Nishiyama et al. (2006) derived a near-infrared extinction relation for stars of the Galactic center using photometry from the SIRIUS survey; they report the following relations: $A_{K_{s}}=1.44\pm0.01E_{H-K_{s}}$ and $A_{K_{s}}=0.494\pm0.006E_{J-K_{s}}$, where $E(H-K_{s})=(H-K_{s})_{\textrm{\scriptsize{obs}}}-(H-K_{s})_{0}$. For the stars presented in this work, the observed $JHK_{s}$ photometry was generally taken from the SIRIUS catalog, except for very bright stars that were saturated ($K_s<8$ mag), in which case we used 2MASS photometry. The intrinsic colors, ${(H-K_{s})}_0$ and ${(J-K_{s})}_0$,  were taken from the literature using comparison stars of similar spectral types. For the WN stars, we used the intrinsic colors from Crowther et al. (2006a, their Table A1).  For the O supergiants we used the synthetic photometry for O supergiants from Martins \& Plez (2006). Using the adopted intrinsic colors, we were able derive two values of $A_{K_{s}}$ using the observed $J-K_{s}$ and $H-K_{s}$ colors, which were then averaged to produce a final value for the $K_{s}$-band extinction. The absolute $K$-band magnitude of each star was derived assuming a distance of 8 kpc to the Galactic center. To calculate the total luminosity of each star, bolometric corrections of comparison stars having similar spectral types were also taken from the same literature sources referenced above. In the interest of a thorough comparison, we not only performed these calculations for new WN and O X-ray sources, but included previously known sources as well. The results are presented in Table 3.   The most luminous stars in the sample are the WN8--9h stars X174656.3 and A7 (Figer et al. 2002) of the Arches cluster (CXOGC J174550.4$-$284919), with $L_{\textrm{\scriptsize{bol}}}=2.9\times10^{6}$ \Lsun and $3.9\times10^{6}$ \Lsun, respectively; the least luminous star is the broad-lined WN7 star X174508.9, which has a luminosity of $L_{\textrm{\scriptsize{bol}}}=2.5\times10^5$ \Lsun. In contrast to WNh stars which are still burning hydrogen, the WN7 stars, such as X174508.9, are in an evolved state of core-He burning. Such stars have already blown off their hydrogen envelope, and as a result, have a significantly smaller radius and lower bolometric luminosity than the less-evolved WNh sources.

The situation is more difficult in the case of the WC stars; we were unable to derive an intrinsic color for these stars, owing to their significant near-infrared excess. Instead, the photometry of X174617.7 was extinction-corrected by adopting the average extinction value for the Quintuplet cluster  ($A_K=2.7$ mag; Figer et al. 1999), since X174617.7 appears to be an outer member of the Quintuplet. For X174519.1, we adopted the extinction value measured for a nearby B2Ia star, S174523.1 from MMM07, which has $A_K=3.2$ mag. Although these extinction approximations will be useful in the analysis of the WC star SEDs, we did not compute bolometric luminosities for them, and they are not included in Table 3. 

\subsection{Infrared Excess}
To determine whether the massive stellar X-ray sources are distinguishable from Galactic center field stars in the infrared, we constructed color--color diagrams. Figure 8 illustrates that almost every star exhibits a significant infrared excess, apparent as an offset from the main reddening locus that is traced by the roughly linear scatter distribution of field stars. In $H-K_s$ versus $J-H$ color space, the WN stars and most of the O supergiants occupy the same space, while the WC9 and DWCL stars appear more separable as a group, owing to their significantly larger $K_s$-band excess. With the inclusion of $\lambda8$ {\micron} photometry, the WN, O, WC9, and DWCL stars are even easier to distinguish as separate groups in $K_s-[8.0]$ versus $J-K_s$ color space. The WN stars and one O supergiant appear to cluster together, while the WC stars, again, exhibit a relatively large excess compared with the WN stars, especially the DWCL stars qF211 and qF231 of the Quintuplet cluster. Three O supergiants exhibit an extremely large excess at $\lambda$8 {\micron}, which likely originates from unresolved, warm gas and dust surrounding these young stars (evidence in support of this hypothesis provided by mid-infrared field images is presented in Section 6.2).

There are several ways in which Wolf-Rayet and O stars can exhibit an infrared excess. Within the ionized winds of hot supergiants, scattering of electrons off of H$^{+}$ and He$^{+}$ ions produces free-free emission, which contributes a significant amount of continuum flux in the infrared, becoming more significant with increasing wavelength as the underlying photospheric continuum becomes fainter with respect to the free-free component (Wright \& Barlow 1975). Thermal emission from hot circumstellar dust ($T=600$--$1300$ K) is another source of infrared excess that is often associated with WC stars (Cohen et al. 1975, Williams et al. 1987). This form of excess emission can exhibit significantly higher flux density than free-free emission. Finally, a large infrared excess may also be detected from massive stars surrounded by unresolved clouds of gas and dust local to the star. 

To further investigate the nature of the infrared excesses observed in our sample, we generated $\lambda1$--10 {\micron} SEDs. For the photometry of the WN and O stars we applied extinction corrections to the photometry in each band according to the extinction ratios in Indebetouw et al. (2005), using the derived values of $A_{K_{s}}$ in Table 3 and the adopted values for the WC stars. To convert the mid-infrared data to flux density units, we applied zero points of 280.9, 179.7, 115.0, and 64.13 Jy for IRAC channels 1, 2, 3, and 4, respectively (IRAC Data Handbook--Table 5.1). In the absence of confusion-induced photometric errors, the near-infrared SIRIUS photometry matches 2MASS very closely; the mean and rms variance of the magnitude differences are 0.009 and 0.015 mag in the \textit{J} band, respectively (Nishiyama et al. 2008). Therefore, we obtained near-infrared fluxes for SIRIUS using the 2MASS zero points of 1594.0, 1024.0, and 666.7 Jy for the $J$, $H$, and $K_{s}$ bands, respectively.  Below, we examine the SEDs of several WN and WC stars to investigate the nature of the infrared excess emission in more detail. 

\subsection{Free-Free Emission from WN/O Stars}
To test the hypothesis that free-free emission is responsible for the infrared excess observed toward the massive stars in our sample, we employed a modified version of the free-free emission model of Cohen et al. (1975). In this model the optically-thin, free-free emission spectrum for an unresolved source may be written as 

\begin{equation}
F_{\nu} (\textrm{Jy}) = 2.3\times10^{-15}~\frac{R_{s}^{3}}{{d}^{2}}\frac{N_{e}N_{i}Z^{2}}{\sqrt{T_{e}}}e^{-C/\lambda T_{e}}
\end{equation}

\noindent{where} $Z$ is the mean electric charge per ion, $N_{e}$ and $N_{i}$ are the respective electron and ion densities in units of cm$^{-3}$, $T_{e}$ is the electron temperature in Kelvins, $R_{s}$ is the size of the emitting region in cm, $d$ is the distance to the source in cm, $\lambda$ is the wavelength in {\micron}, and $C$=14400~{\micron}~K. Free-free emission from evolved massive-star winds typically becomes optically thick in the mid-infrared at some wavelength, $\lambda_{t}$. If this turnover wavelength is observed, one can also solve for the emission measure:

\begin{equation}
N_{e}N_{i}R_{s}=\frac{7.3\times10^{34}\sqrt{T_{e}}}{\lambda_{t}^{3}Z^{2}(1-e^{-C/\lambda_{t}T_{e}})}~\textrm{cm}^{-5}.
\end{equation}

By combining Equations (1) and (2), we constructed model SEDs for several stars: X174516.1(Ofpe/WN9), X174555.3 (WN5--6b), X174656.3 (WN8--9h), and X174712.2 (WN7--8h), all of which have an observable turnover wavelength in their mid-infrared SEDs. For each star we determined the flux density attributable to free-free emission by subtracting off a blackbody continuum. The continuum flux was calculated using the luminosity values in Table 3 and the Stefan--Boltzmann relation $L=4\pi R^{2} \sigma T^{4}$, where $L$ is the bolometric luminosity, $R$ is the stellar radius, and $T$ is the effective temperature. Unfortunately, there is a degeneracy between temperature and radius on the Rayleigh--Jeans portion of the SED, which we are limited to. So, initially we guessed values of $T$ and $R$ that are appropriate for the spectral type of the star being analyzed.  The values were then adjusted until we obtained a satisfactory match to the extinction-corrected near-infrared photometry. Of course, a slight underestimate of $T$ could be compensated by an overestimate of $R$. This uncertainty, however, is unimportant for our purposes because the free-free emission spectrum is not affected by it. Once the resulting Rayleigh--Jeans component was subtracted from the photometry, the remaining free-free flux, described by Equations (1) and (2), was fit using a least squares method. By performing a least squares fit to the excess spectrum, we were ultimately able to derive $T_{e}$, $N_{e}$, $\lambda_{t}$, and $R_{s}$. For the models we assume $N_{e}=N_{i}$ and $Z=1$ for the WNh and Ofpe/WN9 stars, and $Z=2$ and $N_{e}=2N_{i}$ for the hydrogen-deficient WN5--6b star.  

The resulting model parameters are listed in Table 4, and are illustrated graphically in Figure 9. The model curves match the data reasonably well, and the derived values for $T_{e}$, $N_{e}$, and $R_{s}$ are similar to those of the WN stars modeled in Cohen et al. (1975). Reliable errors in our parameter fits are difficult to ascertain; uncertainties in the extinction law and the effect of strong emission lines on the photometry are likely to play a role. Nonetheless, a very precise free-free emission model of the stars in our sample is not the focus here. The goal is to unambiguously demonstrate that the infrared excess emission from the WN stars in Figure 8 is well described by a simple free-free emission model having physical parameters that are typical for such stars. 

\subsection{Hot Dust Emission from WC Stars}
The infrared excess emission from the WC stars in our sample (X174519.1 and 174617.7) has a significantly higher intensity than that of the WN stars, which is apparent in Figure 8.  Figure 10 shows the extinction-corrected SEDs of the new WC9 discoveries X174617.7 and X174519.1, accompanied by the SEDs of the Quintuplet proper member X174614.6 (qF231 from Figer, McLean \& Morris 1999), a putatively single WC9 star HDM13 (Hadfield et al. 2007), and the WC9d star WR 59 for comparison.  For the photometry of the comparison stars HDM13 and WR 59, we applied extinction corrections using the $A_K$ or $A_V$ values from Hadfield et al. (2007) and van der Hucht (2001), respectively. 

For the WC9 stars X174617.7 and X174519.1, we were unable to use our free-free emission model to simultaneously reproduce the near-infrared and mid-infrared flux densities. No sensible combination of the stellar radii, effective temperature, and wind parameters ($T_e$ and $R_s$) resulted in a reproduction of the rise in flux density between the $H$ and $K_s$ bands. Fitting the bright mid-infrared portion of the WC9 SEDs without overestimating the near-infrared portion would require insensibly low values of $T_{e}$. So, even though WC stars are indeed important sources of infrared free-free emission, a stronger additional emission component is required to explain the structure of their infrared SEDs. Thermal emission from hot dust has been observed from many WC9 stars and binaries, and we suspect this is the case for X174617.7 and X174519.1. This hypothesis is strengthened by the following comparison with the other dusty WC stars in Figure 10. qF231 exhibits the largest infrared excess. This star is an X-ray source (X174614.6), and has been classified as a DWCL star (Figer et al. 1999; Moneti et al. 2001). Such stars exhibit nearly featureless near-infrared spectra, as demonstrated in Section 3.5 (Figure 6) with the spectrum of the DWCL star X174645.2 (the \textit{Spitzer}/IRAC photometry of X174645.2 is saturated, so its SED was not included in Figure 10).  By comparison, the infrared excess from the single WC9 star HDM13 is the smallest of the stars in Figure 10, exhibiting characteristics very similar to the WN stars in Figure 9. Thus, the excess from HDM13 can perhaps be attributed to free-free emission alone. The magnitude of the infrared excess in the SEDs of X174519.1 and X174617.7 appears to lie in-between the extremes defined by the SEDs of qF231 and HDM13, but bear the closest resemblance to the WC9d star WR 59. Compared with qF231, the smaller infrared excess of X174519.1 and X174617.7 provides a natural explanation for why the emission lines in their  spectra are only partially diluted, instead of being completely diluted as in the case of the DWCL spectra. Still, the mid-infrared SEDs of these sources exhibit an excess component that rises significantly farther above the RJ continuum than the free-free components of the single WC9 star HDM13 and the WN stars in the previous section. Furthermore, the excess emission peaks between $\lambda$5 and $\lambda$10 {\micron} and is detectable at wavelengths as short as the $J$-band wavelength. Thus, these characteristics can be attributed to the presence of hot dust with temperatures near $\sim1000$ K.  Realistically, the collective excess of these objects is probably due to multiple hot dust components lying at different radii from the central star(s), stellar continuum photons, and free-free emission.  A more detailed modeling of the complex excess emission for these stars is beyond the scope of this paper. We simply intend to demonstrate that hot dust is responsible for the bulk of the mid-infrared excess detected from X174519.1 and X174617.7, and provides a simple explanation for the continuum diluted emission lines observed in their $K$-band spectra. 

WC stars have long been known to be producers of dust if they exist in a binary system. In fact, it has been suggested that all dusty WC stars (WCd) are members of binaries (Williams et al. 2005), in which the dust is produced by the collision of carbon-rich and hydrogen-rich winds from the WC star and an OB companion, respectively. Wind collision also naturally explains the detection of these sources at hard X-ray energies (see Section 6.1.1). For several known WCd stars, the binary hypothesis has been strengthened beyond any reasonable doubt by high-resolution imaging experiments, which have revealed rotating Archimedian spirals of dust surrounding them (e. g., seeTuthill et al. 2006, 2008). Similar structures may be observable, modulo unfortunate projection effects, in association with X174519.1, X174617.7, and X174645.2. 

\section{X-ray Emission}
Table 5 lists the X-ray photometry for all 30 confirmed massive stellar X-ray sources in the Galactic center region, including new and previously identified objects. The data for each star were compiled from the catalog of Muno et al. (2009).  Table 5 includes X-ray source astrometry, total on-source integration time, photon number counts in the hard and soft bands, broadband photon flux, hardness ratios for soft and hard energy bands, the average energy of all detected photons, the derived X-ray luminosity, and the spectral types of their stellar counterparts. Most of the sources have photon fluxes ranging from $F_{\textrm{\scriptsize{X}}}=2\times 10^{-7}$ to $F_{\textrm{\scriptsize{X}}}=7\times 10^{-5}$ cm$^{-2}$ s$^{-1}$. An approximate energy flux can be obtained by multiplying  $F_{\textrm{\scriptsize{X}}}$ by the average photon energy $\langle{E}\rangle$, which yields values ranging from $1.3\times10^{-15}$ to $5.7\times10^{-13}$ erg cm$^{-2}$ s$^{-1}$. 

In order to constrain the nature of the X-ray emission, it is useful to examine the hardness ratio, defined as HR=$(h-s)/(h+s)$, where $h$ and $s$ are the fluxes in the hard and soft energy bands, respectively. The soft color, defined as HR0, is defined by letting $h$ and $s$ be the fluxes in the respective 0.5--2.0 keV and 2.0--3.3 keV energy bands. In Paper I, we used HR0 to select X-ray sources that are likely to be located near the Galactic center, since the softest X-ray photons are highly absorbed by the ISM. For this reason, many of the sources in Table 5 are not detected in the softest 0.5--2.0 keV energy band, which results in them having an HR0 value of 1. The hard color, HR2, is defined by letting $h$ and $s$ be the fluxes in the 4.7--8.0 keV and 3.3--4.7 keV energy bands, respectively. Compared with HR0, the more energetic photons used to calculate HR2 suffer relatively little absorption from intervening gas and dust, so they are more useful for constraining the hard components of the X-ray SEDs. 

Figure 11 is a scatter plot of HR2 versus broadband (0.2--8.0 keV) photon flux for absorbed X-ray sources that are likely to be located near the Galactic center (i.e., those sources whose soft X-ray colors imply a hydrogen absorption column of $N_{\textrm{\scriptsize}}>4\times10^{22}$ cm$^{-2}$).  The majority of the X-ray sources that comprise the ``cloud" of small data points in Figure 11 are likely to be CVs, a hypothesis that is consistent with both the X-ray properties of the sample (Muno et al. 2003, 2009) and the rarity of near-infrared counterparts with $K_s\le15.6$ mag, discussed  in Paper I. By comparison to the field population of presumed CVs, the confirmed massive stars appear to be systematically softer and brighter in X-rays.  To estimate the plasma temperatures of thermal sources, or the power-law indices of non-thermal sources, we constructed models of HR2 versus luminosity for optically-thin thermal plasmas with temperatures of $kT=0.9$, 2.2, 3.4, and 6.8 keV, and power-law emission models for non-thermal sources with photon indices of $-1<\Gamma<3$. The models were constructed for sources lying at a distance of 8 kpc that are absorbed by a hydrogen column of $N_{\textrm{\scriptsize{H}}}=6.0\times10^{22}$ cm$^{-2}$. This value of $N_{\textrm{\scriptsize{H}}}$ is generally adopted as a reasonable average value for X-ray sources near the Galactic center (Muno et al. 2003), and is also consistent with the average extinction of the stars listed in Table 3, according to the relation between $N_{\textrm{\scriptsize{H}}}$ and $A_V$ given in Predehl \& Schmitt (1995).  All models were computed using the \program{XSPEC}\footnote{http://heasarc.gsfc.nasa.gov/docs/xanadu/xspec/} program. The HR2 values of the massive stars imply the presence of plasma components with temperatures of $kT>1$ keV (in most cases $kT>2$ keV), if the emission is thermal, and luminosities in the range $L_{\textrm{\tiny{X}}}\sim10^{32}$--$10^{34}$ erg s$^{-1}$ (0.5--8.0 keV). The X-ray sources with the highest fluxes and hardest colors are the DWCL star X174645.2 (Hyodo et al. 2008) and several of the Arches cluster members, all of which appear to have relatively high luminosities near $L_{\textrm{\tiny{X}}}\sim10^{34}$ erg s$^{-1}$. The softest sources, undetected in the 4.7--8.0 keV band, are the O9I--B0I star X174537.3 and the Quintuplet members qF231 (DWCL star) and qF242 (OBI). The hardest massive stellar X-ray source in the Galactic center is the WN7 star X174508.9, with HR2=0.3, and an average photon energy of 5.3 keV.  We note that a constant value of  $N_{\textrm{\scriptsize{H}}}$ is a gross simplification; more so is the assumption of a single temperature component for the X-ray plasma. Below, we address the effect that these simplifications have on the derived properties of the X-ray-emitting plasma.

The fiducial conversion from photon flux to unabsorbed 0.5--8.0 keV luminosity is such that $10^{34}$ erg s$^{-1}$ equals $6\times10^{-5}$ cm$^{-2}$ s$^{-1}$, assuming $\Gamma=1.5$ or $kT=7$ keV, which are average values for the entire X-ray population. Assuming $kT=2$ keV instead, which is more applicable to our sample of massive stellar X-ray sources, the conversion is such that $10^{34}$ erg s$^{-1}$ equals $4\times10^{-5}$ cm$^{-2}$ s$^{-1}$. However, the uncertainty in these conversion factors is substantial. Indeed, interstellar absorption has extinguished most of the X-ray flux at energies below 1 keV, which is where massive stars emit the bulk of their X-rays (e.g., see Oskinova 2005, Skinner et al. 2002a), so we could be significantly underestimating the total intrinsic X-ray luminosities in the 0.5--8.0 keV range. As a particularly fitting example of the errors imposed by assuming a single temperature plasma, we refer to the X-ray analysis of the source X174536.1 by Mikles et al. (2006). Owing to relatively bright soft and hard X-ray fluxes, the authors were able to fit a detailed thermal plasma model to the X-ray spectrum of this source. Approximating with a single temperature plasma, as we do for our sample as a whole, the authors derived $kT\approx3.5$ keV and $L_{\textrm{\tiny{X}}}=10^{33.8}$ erg s$^{-1}$ (0.5--8.0 keV) for X174536.1, which is  reasonably close to the plasma temperature implied by the HR2 value of this source, and to our derived luminosity of 10$^{33.7}$ erg s$^{-1}$. However, the authors obtained a more accurate spectral fit using a two-temperature model, having a very soft component $kT\approx0.7$ keV and an additional hard component with $kT\approx4.6$ keV, each with respective luminosities of $L_{\textrm{\tiny{X}}}=10^{35.0}$  erg s$^{-1}$ (0.5--2.0 keV) and $10^{33.5}$ erg s$^{-1}$ (2.0--8.0 keV). Thus, the resulting X-ray luminosity difference between the 1-$T$ and 2-$T$ plasma models is a sobering $\approx1.4$ dex.  Clearly, in assigning a single temperature plasma we may be neglecting a significant amount of soft X-ray luminosity, if it lies beneath our flux threshold due to absorption.  Unfortunately, only two of the new detections reported in this work were detected at energies below 2 keV, the brightest of which has only 13 photon counts in that range. Without a significant flux of soft X-rays to subject to a model fit, assuming a single temperature plasma is the best we can currently do. So, we assumed that 2 keV is a reasonable average intrinsic plasma energy and based our luminosity calculations on this assumption. The resulting X-ray luminosities are included in Column 12 of Table 5. As a means of estimating the errors, we compared our values to the results of other authors who were able to derive luminosities for relatively bright, Galactic center X-ray sources based on a modeling of their X-ray spectra, which enable them to solve for $kT$ directly. We did not use the above-mentioned X174536.1 in this exercise because it is a variable X-ray source (see Section 5.1). The comparison is facilitated by Table 6, which includes the three bright WN8--9h X-ray sources from the Arches cluster (Wang, Dong \& Lang 2006), the DWCL star X174645.2 (Hyodo et al. 2008), the WN7 star X174550.6, and the Of star X174528.2 (Muno et al. 2006). The average and standard deviation of the difference in the X-ray luminosity between our results and those of the other authors is 0.25 (0.24) dex. Thus, we assign 3$\sigma$ errors of 0.75 dex for our derived luminosities. The results should still be met with caution, however, since the comparison measurements from the other authors listed in Table 6 also assumed a single temperature plasmas. Finally, using our derived X-ray luminosities and the stellar bolometric luminosities in Table 3, we recovered a relation between these values, such that $L_{\textrm{\tiny{X}}}/L_{\textrm{\tiny{bol}}}=10^{-6.9\pm0.6}$. Figure 12 illustrates this result. 

In conclusion, the main points we intend to convey with X-ray data are that (1) the X-ray photometry of the massive stars as a group is systematically softer and brighter than the field population presumed to be dominated by CVs, (2) each source contains a hard X-ray component, requiring either a thermal plasma with $kT>1$ keV (in most cases $kT>2$ keV), or a power-law source with $-1<\Gamma<3$, (3) the majority of sources have 0.5--8.0 keV luminosities between $L_{\textrm{\tiny{X}}}\sim10^{32}$ and $10^{33}$ erg s$^{-1}$, while several others appear to have higher luminosities closer to $10^{34}$ erg s$^{-1}$, and (4) the $L_{\textrm{\tiny{X}}}/L_{\textrm{\tiny{bol}}}$ ratio for the Galactic center sources has an approximately constant value of $10^{-6.9\pm0.6}$.

\subsection{X-ray Variability}
Variability of the Galactic center X-ray population was addressed in Muno et al. (2009). Variable sources were grouped according to three types of variability:  flux variations between individual observations, separated by day- to year-long time scales (long-term variability), variations that occurred within individual observations (short-term variability), and periodic variations that occurred within an individual observation. Four of the massive stellar X-ray sources near the Galactic center exhibit long-term variability. None of them exhibits short-term or periodic variability. The variable sources include the WN7-8h stars X174536.1 and X174712.2, the Ofpe/WN9 star X174516.1, and the OBI Quintuplet member X174614.5. We reproduce the variability data for these sources from Muno et al. (2008, their Table 6) in our Table 7. The table includes the source record numbers from the original X-ray catalog, the observation ID numbers in which the largest and smallest fluxes were observed, the values of the largest and smallest fluxes, and the ratios of those fluxes. The most variable source among the massive stars is the WN7-8h star X174712.2, which varied by a factor of 3.6 between 2000 March 29 and 2001 July 16. 

A variety of phenomena may give rise to X-ray variability. Colliding-wind binaries with eccentric orbits will increase in X-ray luminosity near periastron. Eclipses of a wind collision zone may also result in variable fluxes if the orbit is close to edge-on. However, colliding-wind binaries may not exhibit detectable X-ray variations, as in the case of WR 25 (Pollock \& Corcoran 2005), which remained stable for a time span of ten years (Raassen et al. 2003). For accreting compact objects in HMXBs, eclipses are also a possible source of X-ray variability, as is an intrinsic variation of accretion rate.  Finally, X-rays from hot spots within an equatorial disk surrounding a massive star (see Section 6.1) may be modulated with the stellar rotation period. The applicability of these models in explaining the characteristics of the X-ray emission of our sample is discussed in the following section.

\section{Discussion}
\subsection{Physical Origin of the X-ray Emission}
The X-ray photometry of the majority of massive stellar X-ray sources near the Galactic center indicates that the emission is from thermal plasmas having temperatures of $kT\gtrsim1$--2 keV, or non-thermal sources with power-law indices of $-1<\Gamma<3$, and X-ray luminosities in the range  $L_{\textrm{\tiny{X}}}=10^{32}$--$10^{34}$ erg s$^{-1}$ (0.5$-$8.0 keV).  Hard-X-ray emission at these luminosities is not a ubiquitous feature of \textit{single} O supergiants or Wolf-Rayet stars; these features are typically regarded as an indication of binarity, since it is well established that hard X-rays can be generated within the shocked interface of the binary components' opposing supersonic winds, or by accretion of stellar wind material onto a compact companion. However, there are extraordinary mechanisms by which some single massive stars can intrinsically generate hard X-rays as well; so, we entertained several interpretations of the hard X-ray emission from our sample.
 
\subsubsection{Intrinsic X-ray Emission from Massive Stars}
The intrinsic X-ray emission from single massive stars throughout the Galaxy is typically thermal and soft, having $kT\approx0.6$ keV (e.g., see Oskinova 2005 and references therein). The X-ray luminosities of O stars follow an observed trend in which they scale with stellar bolometric luminosity as $L_{\textrm{\tiny{X}}}/L_{\textrm{\tiny{bol}}}\approx10^{-7}$ (Long \& White 1980; Berghoefer et al. 1997; Sana et al. 2006; Broos et al. 2007, Albacete Colombo et al. 2007). This trend has been interpreted as a manifestation of the physical link between stellar luminosity, which drives supersonic winds, and the X-ray-emitting shocks that result from those winds via line-driven instabilities (Lucy \& White 1980; Owocki et al. 1988; Feldmeier, Puls, \& Pauldrach 1997). In Section 5, we derived the average and standard deviation of the bolometric to X-ray luminosity ratio for our sample, and obtained $L_{\textrm{\tiny{X}}}/L_{\textrm{\tiny{bol}}}=10^{-6.9\pm0.6}$, which is consistent with the known trend for single massive stars. However, the fact that the $L_{\textrm{\tiny{X}}}/L_{\textrm{\tiny{bol}}}$ relation is obeyed does not necessarily imply that the X-rays are solely generated via radiative shocks within a single stellar wind; colliding-wind binaries have also been shown to obey the $L_{\textrm{\tiny{X}}}/L_{\textrm{\tiny{bol}}}$ relation as well, although many colliding-wind binaries have $L_{\textrm{\tiny{X}}}/L_{\textrm{\tiny{bol}}}$ closer to $10^{-6}$ (e.g., see Oskinova 2005).  Still, the emission of hard X-rays is uncommon property of single massive stars. There is, however, an extraordinary mechanism by which massive stars can genrerate hard X-rays intrinsically. The extraordinary O5.5V star $\theta^{1}$ Orionis C, for example, has an average X-ray luminosity of $L_{\textrm{\tiny{X}}}\sim10^{33}$ erg s$^{-1}$ (0.5--10.0 keV), and a hard spectral component from thermal plasma with $kT\approx3$ keV (Gagn\'{e} et al. 2005). The hard X-ray emission is modulated with the star's 15 day rotation period; so, the hard X-rays have been interpreted as a result of the star's strong, fossil magnetic field ($B\approx1000$ G, Donati et al. 2002), which channels and confines the outflowing stellar wind into an equatorial collision/cooling disk (the magnetic wind-confinement model). The process results in strong shocks that generate plasma temperatures in excess of $10^7$ K, and hence, hard X-rays.  The peculiar Of?p star HD 191612 is another example, having a measured magnetic field of 1500 G (Donati et al. 2006) and an X-ray spectrum containing thermal plasma components with $kT\approx1$--2.5 keV and an X-ray luminosity of 7--$9\times 10^{32}$ erg s $^{-1}$ (Naz{\'e} et al. 2007). HD 191612 exhibits periodic spectroscopic variations (a time-variable equivalent width of the H$\alpha$ line) with a period of 538 days. Walborn et al. (2004) discussed the possibility that the variability is induced by the periastron passage of a binary companion, while Donati et al. (2006) attributed the variations to the star's rotational period and, hence, its strong magnetic field.  Since  $\theta^{1}$ Orionis C and HD 191612 both exhibit periodic changes in their spectra, we should consider the potential to observe the same phenomena in the spectra of the O supergiants of our sample, if they represent similar stellar phenomena. 

The magnetic confinement model has been discussed in attempts to explain the detection of hard X-rays from the nitrogen-rich Wolf-Rayet stars as well, in particular, for WR110, EZ CMa, and the oxygen-rich Wolf-Rayet star WR142, all of which have stellar and X-ray properties that are very similar to our sample and exhibit no evidence for a companion (Skinner et al. 2002a, 2002b; Oskinova et al. 2009).  However, the magnetic wind-confinement model faces difficulties when applied to Wolf-Rayet stars; the wind momenta of Wolf-Rayet stars are over an order-of-magnitude stronger than that of O  stars such as $\theta^{1}$ Orionis C and HD 1919612, and are more difficult to channel effectively. According to Babel \& Montmerle (1997), a stellar wind is confineable if the magnetic energy density exceeds the wind kinetic energy density ($B^2/4 \pi>\rho v^{2}$). Assuming $\rho=\dot{M}/4\pi v r^2$, and $v=0.5v_{\infty}$ at $r=1R_{star}$, this would imply that magnetic fields greater than $\approx$5 kG in strength are necessary to confine the winds of the Wolf-Rayet stars and O supergiants that dominate our sample. Currently, there is no evidence for such strong fields in Wolf-Rayet stars. Strong fields have been considered in the interpretation of hard X-ray emission from the WO star WR 142 (Oskinova et al. 2009); however, this source is intrinsically $\approx$10 times fainter in X-rays than the faintest source in our sample. Thus, we do not currently regard the magnetic wind-confinement scenario as a highly viable model for Galactic center Wolf-Rayet stars, although it may explain the X-ray emission from some of the O stars in our sample. Still, we can not completely rule out the possibility that Wolf-Rayet stars can generate hard X-rays intrinsically by some other exotic mechanism, such as inverse Compton scattering (e.g., see Chen \& White 1991). However, since there are so many Wolf-Rayet and O stars known in the Galactic center that have \textit{not} been detected in X-rays (e.g., in the Arches and Quintuplet clusters), intrinsic hard X-ray generation from our sample, which represents the minority fraction of known massive stars in the Galactic center, would imply that our particular stars are special, having physical properties not ubiquitous to all  Wolf-Rayet stars and O supergiants in the region, which would be puzzling.

\subsubsection{Colliding-Wind Binaries}
The collision of opposing, supersonic winds in massive binaries creates a hot shock ($T\gtrsim10^{7}$ K) that is a potential source of hard thermal X-rays with $kT>1$ keV. It is reasonable to adopt this model in interpreting the hard X-ray emission from our sample, since many known colliding-wind binaries exhibit hard X-rays with luminosities that are comparable to our Galactic center sample, such as WR 147 (WN8+OB; Skinner et al. 2006), WR 25 (WN6ha+O4f ; Raassen et al. 2003; Albacete Colombo et al. 2008),  $\gamma^2$ Velorum (WC8+O7.5; Skinner et al. 2001), V444 Cyg (WN5+O6; Maeda et al. 1999), Cyg OB2 8 (O6If+O5.5III(f); Albacete Colombo et al. 2007), and Plaskett's star (O6I+O7.5I; Linder et al. 2006), all of which have hard spectral components with $kT\approx2$--$4$ keV and $L_{\textrm{\tiny{X}}}=10^{32}$--$10^{34}$ erg s$^{-1}$. With regard to the WC stars of our sample (X174519.1, X174617.7, and X174645.2), the case for colliding winds is particularly strong, since most single WC stars are not detected in X-rays at all (Skinner et al. 2006), and hot dust is present (see Section 4.3). It has been proposed that the production of dust from carbon-rich WC winds requires collision and mixing with a hydrogen-rich companion star's wind (Crowther 2003). While the explanation for hard X-ray emission in the WN/O stars of our sample is not as clear, the colliding-wind hypothesis may be strengthened by the fact that binarity is common in Wolf-Rayet and O stars (Wallace 2007).

If the colliding-wind binary hypothesis is correct, one might expect the signature of a companion star to be evident in the infrared spectra, presented in Section 3. However, the current infrared data are probably of insufficient spectral resolution and signal to noise to distinguish the contributions from individual stellar components of a massive binary. In the case where a Wolf-Rayet star or O supergiant has a fainter OB dwarf companion, the magnitudes may differ by a factor of $\sim$3. With the implied flux ratio of $\approx$16 for the stellar components, H and He emission or absorption lines from the OB companion will be difficult to detect in a moderate-resolution spectrum.  In the case where the binary consists of two similar Wolf-Rayet stars of comparable brightness, the broad emission lines of both components may blend together.  In this case, the spectral lines of individual binary components might only be distinguishable with multi-epoch observations, where radial velocity shifts could betray their presence. Thus, the colliding-wind binary scenario is consistent with the current infrared spectra.  

Assuming our sample is mainly represented by binaries, it is worthwhile to speculate whether their global X-ray properties can be used to constrain the physical parameters of the binary system.  For colliding-wind binaries, the emergent X-ray luminosity from the shocked wind-wind interface is a function of the wind velocities and mass-loss rates of the stellar components, and the binary separation. The exact functional form depends on the nature of the wind collision. For instance, if the wind momenta of two stars in a binary are comparable, then the shock zone will lie somewhere near the center of the space between them, where the opposing winds achieve momentum balance. The collision interface will form a bow shock structure that curves toward the star with the weaker wind. According to Usov (1992; see their Equations (89) and (95)), the total X-ray luminosity in this case can be expressed as the sum of the contributions from two separate layers, one that lies on the inside (concave) surface of the bow shock, composed of material from the wind of the weaker star, and one that lies on the outside (convex) surface of the bow shock, composed primarily of material from the wind of the stronger star. Using mass-loss rates and terminal wind velocities that are typical for the Wolf-Rayet and O stars of our sample ($\dot{\mbox{M}}\sim 10^{-5}$--$10^{-6} {M}_{\odot}$ yr$^{-1}$, $v_{\infty}=1000$--$2000$ km s$^{-1}$; see Section 3), this particular model predicts X-ray luminosities of $L_{\textrm{\tiny{X}}}\sim10^{34}-10^{36}$ ergs s$^{-1}$ for binary separations of $D=0.1$--10 AU. Given our uncertainty in X-ray luminosity, the brightest several sources in our sample could be binaries in which the components have comparable wind momenta. However, the majority of our sources have fainter X-ray luminosities in the range $L_{\textrm{\tiny{X}}}\sim10^{32}-10^{33}$ erg s$^{-1}$. The emergent X-ray luminosity in this particular colliding-wind model has a dependence of $L_{\textrm{\tiny{X}}}\propto D^{-1}$, so, this scenario could result in X-ray luminosities consistent with those of our sample only if the binary separations are very large ($D\gtrsim10$--$500$ AU). This would make the Galactic center sources analogous to the WN8h$+$OB binary WR 147, which has an estimated binary separation of $D\approx400$ AU, and an X-ray luminosity and plasma temperature of $L_{\textrm{\tiny{X}}}(0.5-7~\textrm{keV})\approx6\times10^{32}$ erg s$^{-1}$ and $kT=2-4$ keV, respectively (Skinner et al. 2007). With periods of $\sim$1000 years, binaries like these would not exhibit detectable radial velocity variations. However, at a distance of 8 kpc, they could have apparent separations of $\approx$50 mas, which could be resolvable with adaptive optics on 8--10 m class telescopes. 

Alternatively, Usov (1992) also considers the case where the ram pressure of the stronger stellar wind suppresses the weaker wind of the companion (see their Equation (81)), in which case the companion star is modeled as a hard sphere and the wind collision occurs very near or on its surface. For this particular model, which has a much stronger dependence on the binary separation ($D^{-4}$), the expected X-ray emissivity of the shock zone yields lower values of X-ray luminosity in the range of $L_{\textrm{\tiny{X}}}\sim10^{32}$--$10^{33}$ ergs s$^{-1}$. Thus, if this particular model is applicable, then perhaps the Galactic center sources are binaries where the components have significantly different wind strengths. This model has been considered to explain the X-ray emission of the WN5--6 star WR 110 (Skinner et al. 2002a), which has $L_{\textrm{\tiny{X}}}(0.5-7~\textrm{keV})\sim10^{32}$ erg s$^{-1}$ and a hard spectral component with $kT\gtrsim3$ keV. 

It is interesting that the most X-ray-luminous stars in our sample, with $L_{\textrm{\tiny{X}}}\sim10^{34}$ erg s$^{-1}$, are mainly comprised of objects within the Arches cluster. These particular stars could be binaries that have relatively small separations. Frequent stellar encounters within dense stellar clusters will tend to harden close binaries while `ionizing' wide ones; so, it is reasonable to speculate that the relatively high X-ray luminosity of the Arches X-ray sources is the result of their relatively small orbital separations that were imposed by the dense cluster environment. If so, radial velocity variations will occur on shorter timescales than for the rest of our sample and may exhibit a larger amplitude in the radial velocity variations, similar to that of the WN6h$+$O binary NGC 3603-A1, which lies near the center of the starburst cluster NGC 3603 and exhibits radial velocity amplitudes of $\approx400$ km s$^{-1}$ (Schnurr et al. 2008).   
 
\subsubsection{Supergiant High-Mass X-Ray Binaries}
We must also consider the possibility that wind-accreting neutron stars or black holes are responsible for the X-ray emission from some of the massive stellar X-ray sources near the Galactic center, making them supergiant X-ray binaries (SGXBs). Although there is no reason to expect such objects to follow the observed trend of $L_{\textrm{\tiny{X}}}/L_{\textrm{\tiny{bol}}}=10^{-7}$, there are several quiescent SGXBs whose $L_{\textrm{\tiny{X}}}/L_{\textrm{\tiny{bol}}}$ ratios that are consistent with this trend. So, the wind-accretion model still deserves consideration.

According to Pfahl, Rappaport \& Podsiadlowski (2002), the time-averaged 1--10 keV accretion luminosity of a wind-accreting neutron star is given by
\begin{equation}
\frac{{\langle L_{\textrm{\scriptsize{X}}}\rangle}}{10^{33}~\textrm{erg s}^{-1}} \sim \eta \left( \frac{ \dot{M}_w}{10^{-8}~\Msun~\textrm{yr}^{-1}} \right) \left( \frac{a}{0.5~\textrm{AU}} \right)^{-2} \left( \frac{v_w}{1000~\textrm{km s}^{-1}} \right)^{-4} (1-e^{2})^{-1/2},
\end{equation}
\noindent{where} $\eta$ is the conversion efficiency of gravitational energy into radiation,  $\dot{M}_w$ and $v_w$ are the mass-loss rate and wind velocity of the donor star, respectively, $a$ is the orbital separation in AU, and $e$ is the orbital eccentricity of the system. This formula is based on the Bondi--Hoyle--Lyttleton accretion model (Hoyle \& Lyttleton 1941; Bondi \& Hoyle 1944). For the massive stars of our sample, most of which have $\dot M \sim 10^{-6}$--$10^{-5}~\Msun$ yr$^{-1}$ and $v\approx2000$ km s$^{-1}$, Equation (3) implies that the emergent X-ray luminosity from a wind-accreting SGXB with these wind parameters is $L_{\textrm{\scriptsize{X}}}\sim10^{31}$--$10^{34}$ erg s$^{-1}$ for binary separations ranging from 0.5 AU to several AU, assuming a radiative efficiency of 1. 

Currently, there are very few physical examples of SGXBs with X-ray luminosities of  $L_{\textrm{\scriptsize{X}}}\sim10^{34}$ erg s$^{-1}$ or lower, presumably because X-ray surveys of the Galactic plane are biased toward bright SGXBs (e.~g., Roche-lobe overflow systems with $L_{\textrm{\tiny{X}}}\sim10^{36}$ erg s$^{-1}$; Liu et al. 2000). However, a new class of SGXB has recently been unveiled in obscured regions of the Galactic plane by hard X-ray surveys, which may provide some insight into X-ray production in low-accretion-rate systems: the superfast X-ray transients (SFXT; Negueruela et al. 2006a; Sguera et al. 2006) are SGXBs containing hot supergiant primary stars. These systems undergo very brief X-ray outbursts, in which the X-ray flux increases by a factor of 10$^4$--10$^5$ for durations on the order of hours, in-between much longer states of quiescence ($L_{\textrm{\scriptsize{X}}}\sim10^{33}$ erg s$^{-1}$) that last on the order of weeks to a month. Approximately 10 such objects are currently known, many of which contain X-ray pulsars. The outbursts are thought to occur when a neutron star in an eccentric orbit nears periastron, and plunges into dense inhomogeneities within the hot supergiant's wind. The long quiescent periods ensue as the neutron star travels further from the primary; it is this particular state of SFXTs which we intend to compare with our sample of Galactic center X-ray sources. For example, during their quiescence states the SFXTs XTE J1739$-$302 and IGR J11215$-$5952 exhibit hard power-law spectra having $\Gamma \approx 1$--2 and X-ray luminosities of $L_{\textrm{\scriptsize{X}}}\approx(3$--4$)\times10^{33}$ erg s$^{-1}$ within the 0.1--10 keV energy range (Sidoli et al. 2008; Romano et al. 2007, 2009). If these particular values of $\Gamma$ and $L_{\textrm{\scriptsize{X}}}$ for XTE J1739$-$302 and IGR J11215$-$5952 were plotted in Figure 11, they would be practically indistinguishable from the stars of our sample.  Thus, the X-ray properties of the Galactic center sources are consistent with the quiescent states of SFXTs. Furthermore, quiescent X-ray emission from an SFXT is also consistent with Equation (3): IGR J11215$-$5952, which contains the B1Ia star HD 306414, exhibits evidence for an orbital period of $P\approx165$ days (Romano et al. 2009). B1Ia stars have $M\approx40$--50 \Msun,  $\dot{M}_w\sim10^{-6} \Msun$ yr$^{-1}$, and $v_w\sim1000$ km s$^{-1}$ (Crowther et al. 2006a, 2006b). So, Kepler's law implies that this system should have a separation of $a\approx2$ AU. Thus, Equation (3) implies an X-ray luminosity of  $L_{\textrm{\scriptsize{X}}}\approx 6\times10^{33}$ erg s$^{-1}$ for this system, which is a close approximation to this source's lowest observed quiescent luminosity of $L_{\textrm{\scriptsize{X}}}=4\times10^{33}$ erg s$^{-1}$ (Romano et al. 2007). 

However, although there has been moderate X-ray variability observed for the several sources listed in Table 7, there is no evidence that any of the stars of our sample underwent eruptions or outbursts that are on the level known to occur in SFXTs. Thus, we are not suggesting that the stars in our sample are SFXTs; rather, we are examining the X-ray properties of quiescent SFXTs to provide some insight into the nature of faint X-ray emission from SGXBs with relatively wide binary separations. Indeed, SGXBs with wider and more circular orbits could exhibit the quiescent X-ray properties of SFXTs without the periodic outbursts.  

In conclusion, the current data are consistent with both the colliding-wind and wind-accretion hypotheses. The most effective method of discriminating between these two possibilities is to derive a constraint on the system mass functions. Colliding-wind binaries will exhibit variable doppler shifts in their spectra on the order of $\approx10$--500 km s$^{-1}$, modulo unfavorable binary inclinations or large binary separations, while the doppler signature of the gravitational tug of a wind-accreting neutron star ($M\approx2\Msun$) in the spectrum of a supergiant donor star would be more challenging to detect. We plan to perform such experiments in the near future.

\subsection{Spatial Distribution, Environments, and Implications for Star Formation}
Mid-infrared images from $Spitzer$/IRAC of the fields containing the massive stellar X-ray sources are presented in Figures 13--15. For Figures 13 and 14, three-color images were assembled using data from IRAC channels 1 ($\lambda$3.6 {\micron}), 3 ($\lambda$5.8 {\micron}), and 4 ($\lambda$8.0 {\micron}), and are displayed with a histogram-equalization contrast stretch to enhance subtle structure in surrounding nebulosity. Each image is 144\arcsec~on a side, which corresponds to a projected physical distance of $\approx$5.6 pc, assuming a distance of 8 kpc to the Galactic center. The images bear the name of the X-ray source that lies at the field center as well as alphabetical labels that show the location of the field in the wide-field $\lambda$8.0 {\micron} image presented in Figure 15. 

Several fields in Figures 13--15 contain multiple massive stellar X-ray sources or additional, known massive stars: field D of the WC9d star X174519.1 also contains the Ofpe/WN9 star X174516.1 on the western side of the image and the B2 Ia star from MMM07 on the eastern side; field F of the O4--6I star X174532.7 (the central source of the H1 nebula) contains the Of star X174528.6 on the western side of the image and the WN7--8h star X174536.1 (Mikles et al. 2006) on the eastern side; field G of the O9I--B0I star X174537.3 contains the OBe star H5 (Cotera et al. 1999) to the northeast of the field center and the WC8--9 star WR 101p (Homeier et al. 2003) to the far northeast; finally, field K of the WC9d star X174517.7 also contains the central portion of the Quintuplet cluster toward the north-west of the image, which is dominated by the infrared emission of very bright DWCL Quintuplet proper members, including X174514.6 (qF231) and X174515.8 (qF211). The LBV Pistol Star and its spherical ejection nebula are also apparent in field K.  

Several massive stellar X-ray sources appear to have diffuse mid-infrared nebulae associated with them. In particular, each of the three O supergiants that exhibit the largest $\lambda8.0$ {\micron} excess in Figure 8 (X174537.3, X174628.2, and X174528.6) appears to be surrounded by such nebulae, which is extended in the case of 174628.2, and more compact in X174537.3, and  X174528.6. The excess point-source emission may be natal gas and dust in the process of being heated and cleared away by the strong radiation fields and winds of the supergiants.  Although we did not present the SEDs of these stars in Section 5, we did examine them, and determined that they appear very similar to the bow-shock source IRS8 near the central parsec, which although unresolved by \textit{Spitzer}, was clearly resolved by Geballe et al. (2006) via adaptive-optics imaging. Hence, we speculate that the bright $\lambda8.0$ {\micron} excess from X174537.3, X174628.2 and X174528.6 may be unresolved dust emission from bow-shocks or heated dust structures that are similar to that surrounding IRS8. The WN7-8h stars X174516.7 and X174712.2 also appear to be associated with bright mid-infrared nebulae, although in these cases the nebulae are well resolved from the central stellar point source.  The O9I--B0I star X174502.8 and the WN7--8h star X174532.7 also appear to be having strong impact on their surroundings, as evidenced by the bow-shock-like or shell-like morphology of nearby nebulae. 

The origin of the massive stellar X-ray sources near the Galactic center is currently unclear. With the exception of several rather obvious associations with the Arches and Quintuplet clusters, it is possible that the relatively isolated stars near the clusters (e.~g., X174555.3 and X174517.0) originated within them but were dynamically ejected via gravitational interactions with other massive stars and binaries within the cluster, or were ejected as the result of a companion supernova. Indeed, a noteworthy fraction of O stars ($\approx$10\%--30\%) in the Galactic disk are known to be runaways, $\approx$10\% or more of which are binaries (Gies \& Bolton 1986; Gies 1987). Runaway Wolf-Rayet+O binaries are also known to exist, such as WR 21 and WR 22 in Carina (Moffat et al. 1998). Thus, it is likely that some massive stars and binaries have been dynamically ejected from the Arches and Quintuplet clusters over their respective lifetimes of 1--2 Myr (Figer et al. 2002) and 4--6 Myr (Figer, McLean \& Morris 1999). It is also tempting to speculate on the ejection scenario for the DWCL star X174645.2 (Figure 15, subfield M), owing to its apparent solitude and relatively high latitude above the Galactic plane, compared with the other massive stellar X-ray sources. Unfortunately, there is no obvious potential point of origin for this evolved star, so its birthplace remains elusive.  Proper motion measurements performed with adaptive optics, and high-resolution spectra, would be required to obtain the kinematic information necessary to place meaningful constraints on the origins of the isolated massive stars.

Alternatively, some of the stars from our sample may be the products of a mode of isolated massive star formation, operating in tandem with the formation of dense stellar clusters. The relatively loose aggregations of early-O and WNh stars near the H1--H8 H {\sc ii} regions (Figure 13, fields E and F) and Sagittarius B, and the more evolved WN and WC stars southwest of Sagittarius A West might be the products of such a formation mode. However, it is also possible that some of these stars are members of stellar clusters less extreme in mass and density than the Arches and Quintuplet, which have gone unnoticed owing to confusion with the dense stellar background. Indeed, clusters near the Galactic center are subject to strong tidal forces, causing them to dissolve on time scales of $\sim10$ Myr (Kim et al. 1999), while they may fade beyond detectability as a surface density enhancement within only a few Myr (Portegies-Zwart et al. 2001). If such hidden clusters exist, narrow-band imaging in the vicinity of the newly discovered stars of our sample could betray their presence with the detection of additional, coeval emission-line stars.  

Although the origin of the massive stellar X-ray sources reported in this work is puzzling, it is clear that significant massive star formation has been occurring in the Central Molecular Zone, outside of the well known clusters, for several Myr, including the formation products of the Sagittarius B molecular cloud complex and the newly discovered regions southwest of Sagittarius A West. The \textit{Chandra} observations of the Arches and Quintuplet clusters indicate that only a small fraction of massive stars in the Galactic center are detectable as hard X-ray sources. Indeed, the Arches and Quintuplet contain over 100 massive stars each ($M\gtrsim20\Msun$), yet only several ($\approx5$\%) are detected by \textit{Chandra} as X-ray sources. If roughly the same fraction of massive stars is detectable in X-rays throughout the entire Central Molecular Zone, we speculate that the existence of $\approx$20 massive stellar X-ray sources outside of the stellar clusters implies the presence of $\approx$400 additional massive stars that have yet to be identified.  It is noteworthy that the total inferred Lyman-$\alpha$ photon production rate that is required to explain the far-IR luminosity emerging from the central half-kiloparsec ($\sim10^{52}$ photons s$^{-1}$) is about twice the amount generated by the three clusters collectively (Cox \& Laureijs 1989, Figer et al. 2004); the several hundred massive stars implied to exist by the results of this work would account for the current deficit. However, dense clusters may dynamically produce more hard X-ray sources than less dense associations, in which case the number of massive X-ray-emitting stars detected outside of the clusters would imply an even larger population of undetected massive stars.

\section{Summary and Concluding Remarks}
We have confirmed 16 new massive stars in the Galactic center through infrared spectroscopy of near-infrared counterparts to \textit{Chandra} X-ray sources. The spectral types include early- and late-O supergiants, as well as WN and WC Wolf-Rayet stars. All of the stars exhibit infrared excess, which is attributable to free-free emission from ionized stellar winds, supplemented by thermal emission from hot dust in the case of several WC stars. For several cases we modeled the free-free emission and derived stellar wind parameters, producing results that are in excellent agreement with previous studies of similar stars.

The X-ray photometry of the sources exhibit significant flux at energies above $E>2$ keV, which we regard as an indication of binarity since this property is not typical of single massive stars. Our favor currently leans toward the colliding-wind binary hypothesis for the majority of our sample, mostly owing to similarities in their X-ray properties with known objects from the literature, and the commonality of such objects among massive-star associations in the Galaxy. However, the current data are also consistent with the presence of accretion-powered SGXBs in our sample. It should be noted, however, that without direct evidence for companions, intrinsic hard X-ray generation from single stars, either via magnetically confied winds or some other mechanism, cannot be completely ruled out. Only a constraint on system mass functions via photometric and radial velocity monitoring will allow further discrimination between these alternate hypotheses, and we plan to conduct such experiments in the near future. 

The newly discovered Wolf-Rayet stars and O supergiants bring the total number of massive stellar X-ray sources to 31 (30 if we regard the X-ray detection of X174502.8 as spurious; see Section 3.1), including those within the Arches and Quintuplet clusters. The detection of $\approx20$ such sources outside of the clusters suggests there may be a large population of several hundred massive stars that are currently unidentified in the Galactic center region, since we expect no more than $\approx$5\% of massive stars in the Galactic center to be detectable in X-rays within our current sensitivity limit. If such a large population exists, it would account for the missing Lyman-$\alpha$ photon flux required to explain the far-IR luminosity emerging from the central half-kiloparsec. 

Finally, in Paper I we estimated that there are likely to be $\approx$100--300 absorbed X-ray sources having real infrared counterparts with $K_s<15.6$ mag. Expanding the spectroscopic search to include fainter infrared matches to X-ray sources will undoubtedly reveal more massive stellar counterparts in the Galactic center, including additional Wolf-Rayet and O stars, and HMXBs. Such discoveries will elucidate the origin of these stars and their overall distribution throughout the Galaxy's Central Molecular Zone. The next generation of near-infrared multi-object spectrographs on 8--10 m telescopes, such as FLAMINGOS2 on Gemini-South (Eikenberry et al. 2008) and MOSFIRE on Keck (McLean et al. 2008), will be well suited for this task.

\begin{acknowledgments}
This research was based upon observations with the \textit{Chandra X-ray Observatory}. Support  was provided by the National Science Foundation (grant AST-0406816).  J. M. thanks Fabrice Martins for providing model stellar spectra of the Arches stars for comparison, and comments on the spectral classification of the massive stars presented in this work. We also thank Shogo Nishiyama and Tetsuya Nagata for providing near-infrared photometry from the SIRIUS survey. 

\end{acknowledgments}

\begin{deluxetable}{lcccccc}
\tablecolumns{6}
\tablewidth{0pc}
\tabletypesize{\scriptsize}
\tablecaption{Infrared Spectroscopic Observations of Counterparts to X-Ray Sources}
\tablehead{
\colhead{Associated X-Ray Source} & \colhead{Observation Date (UT)} & \colhead{Telescope/Instrument} & \colhead{${\lambda}/\delta{\lambda}$} &\colhead{Seeing}& \colhead{Sky Cond.}
}
\startdata
CXOGC J174550.6$-$285919   & 2002 May 8 11:22   & IRTF/SpeX      & 2000 &\nodata &clear\\ 
CXOGC J174617.7$-$285007   & 2006 Jul 9 11:06   & AAT/IRIS2       & 2400 &1.2\arcsec &clear\\ 
CXOGC J174555.3$-$285126   & 2006 Jul 20 08:22   & KeckII/NIRC2    & 2200 &0\farcs7--0\farcs9  &clear\\ 
CXOGC J174617.0$-$285131   & 2006 Jul 20 08:48   & KeckII/NIRC2    & 2200 & 0\farcs7--0\farcs9&clear\\ 
CXOGC J174656.3$-$283232   & 2007 May 8 13:53   & IRTF/SpeX      & 250 &0\farcs8--1\arcsec & thin clouds\\   
CXOGC J174711.4$-$283006   & 2007 May 9 11:28   & IRTF/SpeX      & 250   &0\farcs6--0\farcs9& thin clouds\\ 
CXOGC J174703.1$-$283119   & 2007 Jun 15 08:26   & UKIRT/UIST    & 2133 &0\farcs4--0\farcs5&thin clouds\\ 
CXOGC J174725.3$-$282523   & 2007 Jun 15 09:02   & UKIRT/UIST    & 2133&0\farcs3--0\farcs5&thin clouds\\  
CXOGC J174628.2$-$283920   & 2007 Jun 16 09:45   & UKIRT/UIST    & 2133 &0\farcs5--0\farcs6&clear\\ 
CXOGC J174713.0$-$282709   & 2007 Jun 28 06:55   & UKIRT/UIST    & 2133 &0\farcs9& thin clouds\\ 
CXOGC J174519.1$-$290321   & 2007 Jun 28 07:50   & UKIRT/UIST    & 2133 &0\farcs9& thin clouds\\ 
CXOGC J174537.3$-$285354   & 2007 Jul 20 06:52   & UKIRT/UIST    & 2133 &0\farcs7--0\farcs1&clear\\ 
CXOGC J174532.7$-$285617   & 2007 Jul 22 06:38   & UKIRT/UIST    & 2133 &0\farcs7--0\farcs8 & thin clouds \\ 
CXOGC J174712.2$-$283121   & 2007 Jul 22 07:36   & UKIRT/UIST    & 2133 &0\farcs7--0\farcs8&thin clouds\\ 
CXOGC J174616.6$-$284909   & 2008 May 14 12:51   & AAT/IRIS2        & 2400&0\farcs8 &clear \\ 
CXOGC J174516.1$-$290315   & 2008 May 16 16:07   & AAT/IRIS2        & 2400 &1.3\arcsec &clear\\ 
CXOGC J174508.9$-$291218   & 2008 Jun 16 04:01   & SOAR/OSIRIS & 3000 &0\farcs8&30--50\% cover\\ 
CXOGC J174502.8$-$290859   & 2008 Jun 16 07:29   & SOAR/OSIRIS & 3000 &0\farcs8&30--50\%  cover\\ 
CXOGC J174516.7$-$285824   & 2008 Jun 17 05:33   & SOAR/OSIRIS & 3000 &1\arcsec&30--50\% cover \\ 
CXOGC J174522.6$-$285844    & 2009 Jun 14 01:34 & SOAR/OSIRIS & 1200 &1\arcsec & thin clouds \\ 
 CXOGC J174645.2$-$281547   & 2009 Aug 05 08:45   & IRTF/SpeX & 1200 &0\farcs6 &clear 

\enddata
\end{deluxetable}

\setlength{\tabcolsep}{0.06in}
\renewcommand{\arraystretch}{1.0}
\begin{landscape}
\begin{deluxetable}{cccccccccc}
\tablecolumns{11}
\tablewidth{0pc}
\tabletypesize{\scriptsize}
\tablecaption{Infrared Photometry of Massive Stellar X-ray Sources}
\tablehead{
 \colhead{X-Ray Source} & \colhead{R.A.$_{\textrm{\tiny{IR}}}$} & \colhead{Decl.$_{\textrm{\tiny{IR}}}$} & \colhead{$J$} & \colhead{$H$} & \colhead{$K_s$} &  \colhead{$M_{3.6}$}  & \colhead{$M_{4.5}$}  & \colhead{$M_{5.8}$} &\colhead{$M_{8.0}$} \\ [2pt]
 \colhead{(CXOGC J)} & \multicolumn{2}{c}{(deg, J2000)} & \colhead{(mag)} & \colhead{(mag)} & \colhead{(mag)} & \colhead{(mag)} & \colhead{(mag)} & \colhead{(mag)} & \colhead{(mag)}}
\startdata
 174502.8$-$290859\tablenotemark{a} & 266.26195 & $-29.14986$ & $ 13.93 \pm 0.05 $ &$ 11.43 \pm 0.02 $&$ 9.88 \pm 0.01$ & $8.65\pm0.01 $ &$8.30\pm0.01$&$8.31\pm0.02$& \nodata \\           

 174508.9$-$291218 &  266.28733 & $-29.20495$ & $15.12 \pm  0.03$ & $12.61 \pm  0.02$ & $11.08 \pm  0.04$ & $8.88 \pm  0.01$ & $8.42 \pm  0.01$ & $8.15 \pm  0.02$ &  $8.06 \pm  0.05$ \\ 

 174516.1$-$290315 & 266.31744 & $-29.05430$ & $11.49 \pm  0.02$ & $ 9.17 \pm  0.03$ & $ 7.89 \pm  0.03$   &$ 6.76 \pm  0.00$ & $ 6.28 \pm  0.01$ & $ 6.10 \pm  0.01$ & $ 5.76 \pm  0.02$\\ 

174516.7$-$285824 &  266.31969 & $-28.97364$ & $16.67 \pm  0.05$ & $13.08 \pm  0.03$ & $11.09 \pm  0.02$ & $ 9.40 \pm  0.01$ & $ 8.83 \pm  0.01$ & $ 8.68 \pm  0.03$ & $ 7.38 \pm  0.03$ \\

174519.1$-$290321 & 266.32974 & $-29.05609$ & 17.10 &  $13.30 \pm  0.05$ & $10.40 \pm  0.05$ & $ 7.64 \pm  0.01$ & $ 6.78 \pm  0.01$ & $ 6.55 \pm  0.01$ & $ 6.89 \pm  0.03$\\ 

174522.6$-$285844 & 266.34453 & $-28.97895$  & \nodata & $15.21\pm0.06$ & $12.22\pm0.03$ & $10.01\pm0.06$ & $9.13\pm0.08$ & $8.92\pm0.04$ & \nodata \\   

 174528.6$-$285605 & 266.36922 & $-28.93476$& $14.46 \pm  0.02$ & $11.46 \pm  0.02$ & $ 9.72 \pm  0.03$ &$ 6.58\pm0.01 $&$5.21\pm0.01 $&$ 3.69\pm0.01$&$1.43\pm0.01$ \\ 

174532.7$-$285617 & 266.38652 & $-28.93797$ &    $14.67 \pm  0.02$ & $12.13 \pm  0.01$ & $10.72 \pm  0.01$ & \nodata & \nodata & \nodata & \nodata  \\ 

 174536.1$-$285638 & 266.40056 & $-28.94405$ & $15.55 \pm  0.03$ & $12.26 \pm  0.01$ & $10.42 \pm  0.02$ & $ 8.82 \pm  0.01$ & $ 8.26 \pm  0.01$ & $ 7.97 \pm  0.02$ & $ 7.82 \pm  0.06$\\ 

 174537.3$-$285354\tablenotemark{b}  & 266.40538 & $-28.89827$ & $15.75\pm0.02 $&$12.80\pm0.01$ & $11.23\pm0.01$&   $ 9.03 \pm  0.01$ & $ 8.18 \pm  0.01$ & $ 6.56 \pm  0.01$ & $ 4.50 \pm  0.01$   \\ 

174549.7$-$284925\tablenotemark{c}  & 266.45734  & $-$28.82389 &  \nodata                & $14.26\pm0.06$ & $12.16\pm0.04$ & \nodata & \nodata & \nodata & \nodata \\ 
174550.2$-$284911  & 266.45941  & $-$28.81994 &  $15.24\pm0.02$ &  $12.16\pm0.02$ & $10.47\pm0.01$ &$9.21\pm0.01$ & $8.81\pm 0.01$& $8.82\pm0.04$ & \null \\ 
 174550.4$-$284922 & 266.46007 & $-$28.82287 & $14.94\pm0.03$ & $11.64\pm0.02$ & $9.87\pm0.03$ & \nodata & \nodata & \nodata & \nodata \\ 
174550.4$-$284919 &  266.46022 & $-$28.82212 & $14.48\pm0.05$ & $11.45\pm0.02$ & $9.56\pm0.06$ & \nodata & \nodata & \nodata & \nodata \\ 

 174550.6$-$285919  & 266.46087 & $-28.98878$ & $15.44 \pm  0.02$ & $12.48 \pm  0.01$ & $10.85 \pm  0.01$ & $ 9.63 \pm  0.01$ & $ 9.00 \pm  0.01$ & $ 8.63 \pm  0.02$ & $ 7.90 \pm  0.04$ \\ 

 174555.3$-$285126 & 266.48067 & $-28.85738$ & $15.43\pm0.19$& $ 12.53\pm0.04 $ &$10.97 \pm 0.07$ & $ 9.43 \pm  0.01$ & $ 8.70\pm0.01$ & $8.54 \pm 0.03$ &  $8.35\pm0.09$ \\ 

174614.5$-$284937 &  266.56044  & $-28.82688$ & $14.9\pm0.2$ & $12.6\pm0.2$ & $10.7\pm0.2$&\nodata & \nodata & \nodata & \nodata \\ 

174614.6$-$284940 & 266.56131 & $-28.82805$ & 13.55 & $ 10.69 \pm  0.11$ & $7.29\pm0.02$ & $3.16\pm0.01$& $1.84\pm0.01$& $0.81\pm0.01$ & $0.32\pm0.01$  \\

174615.1$-$284932 & 266.56318 & $-$28.82583 &$14.10\pm0.02$ & $11.10\pm0.02$ & $10.59\pm0.20$ &  \nodata & \nodata & \nodata & \nodata \\ 

 174615.8$-$284945 & 266.56611 & $-28.82934$ & $14.90\pm0.08$ & $10.41\pm0.04$ & $ 7.24 \pm 0.03$ & $3.53\pm0.01$ & $3.20\pm0.01$ & $1.38\pm0.01$ & $0.98\pm0.01$ \\ 

174616.6$-$284909 & 266.56946 & $-29.81931$ &  $16.53 \pm  0.03$ & $12.99\pm0.01$ & $11.08\pm0.02$ & $9.42\pm0.07$ & \nodata & \nodata & \nodata \\  

174617.0$-$285131 & 266.57119 & $-28.85871$ &$14.98 \pm0.03$&$12.12 \pm0.02 $&$10.49\pm0.02$ & $9.01\pm0.01$ & $8.66\pm0.01$ & $8.40\pm0.02$ & $8.50\pm0.05$ \\

 174617.7$-$285007 & 266.57430 & $-28.83541$ & $13.99 \pm0.05$ & $10.37 \pm  0.06$ & $ 7.84 \pm0.03$ &$5.52 \pm  0.01$ & $ 5.02\pm0.01$ & $3.92 \pm 0.01$ & $ 3.65 \pm  0.01$ \\ 

 174645.2$-$281547 & 266.68853 & $-28.26324$ &  15.39  & $9.96 \pm  0.03$ & 7.18  &saturated & saturated & saturated & 0.63$\tablenotemark{d}$  \\ 

 174628.2$-$283920 & 266.61793 & $-28.65579$ & $16.99 \pm  0.07$ & $13.36 \pm  0.03$ & $11.49 \pm  0.04$  &\nodata & \nodata & \nodata & \nodata \\ 

174656.3$-$283232 & 266.73468 & $-28.54234$ & \nodata &  $13.74 \pm  0.05$ & $11.24 \pm  0.02$ & $ 9.50 \pm  0.01$ & $ 8.42 \pm  0.01$ & $ 8.11 \pm  0.02$ & $ 8.25 \pm  0.05$ \\   

 174703.1$-$283119 & 266.76307 & $-28.52225$ &  $16.23 \pm 0.03$ & $13.03 \pm 0.01$ & $11.27 \pm 0.01$  &$7.52\pm0.01$&$6.31\pm0.01 $&$5.00\pm0.01 $&$3.45\pm0.01 $   \\  

 174711.4$-$283006 &  266.79782 & $-28.50194$ &  $16.56 \pm  0.06$ & $12.72 \pm  0.02$ & $10.54 \pm  0.01$ & $9.29\pm0.01$ &$ 8.53\pm0.01 $& $8.33\pm0.02$ & $8.51\pm0.06 $\\  

174712.2$-$283121 & 266.80104 & $-28.52267$ &    $17.06 \pm  0.07$ & $13.07 \pm  0.01$ & $10.78 \pm  0.02$ & $ 9.01 \pm  0.02$ & $ 8.21 \pm  0.01$ & $ 7.42 \pm  0.03$ & \nodata\\ 

 174713.0$-$282709 & 266.80423 & $-28.45251$ & \nodata &  $14.22 \pm  0.01$ & $11.86 \pm  0.01$  & $10.57 \pm  0.01$ & $ 9.65 \pm  0.01$ & $ 9.34 \pm  0.03$ & \nodata \\ 

 174725.3$-$282523 & 266.85567 & $-28.42304$ &\nodata &  $13.37 \pm  0.01$ & $11.30 \pm  0.04$ & $10.24 \pm  0.01$ & $ 9.59 \pm  0.01$ &  \nodata   &\nodata   
\enddata
\tablecomments{All known massive stars in the Galactic center with X-ray counterparts are listed, in addition to those discovered in this work. IR source positions and \textit{JH$K_{s}$} photometry are taken from Paper I. The $\lambda$3.6--$\lambda$8.0 {\micron} photometry is from the \textit{Spitzer}/IRAC point-source catalog of Ram{\'{\i}}rez et al. (2008). Uncertainties are included where available. }
\tablenotetext{a}{X-ray source considered spurious and not included in main X-ray catalog of Muno et al. (2009). This may be an X-ray source near our limit of sensitivity (see text). }
\tablenotetext{b}{This source did not make the final candidate counterpart list in Paper I; it has an IR positional error greater than 0.1\arcsec. We suspect this large error is a result of hot dust emission contribution to the $K_s$-band, since this source also has a very bright $\lambda$8 {\micron} counterpart.}
\tablenotetext{c}{$JHK$ photometry for this source from Figer et al. (1999). Note that in this case a $K$-band measurement is presented, rather than the $K_s$ band.}
\tablenotetext{d}{$\lambda$8 {\micron} data from the MSX Point Source Catalog, assuming the zero-point flux of 58.4 Jy from Cohen et al. (2000).}
\end{deluxetable}
\end{landscape}

\begin{landscape}
\begin{deluxetable}{llrccccccccccc}
\setlength{\tabcolsep}{0.06in}
\renewcommand{\arraystretch}{1.0}
\tabletypesize{\scriptsize}
\tablecolumns{13}
\tablewidth{0pc}
\tablecaption{Spectral Types and Absolute Photometry for WN and O Supergiant X-ray Sources}
\tablehead{\colhead{X-Ray Source} & \colhead{Spectral}& \colhead{$K_{s}$} & \colhead{$(J-K_{s})_{0}$}& \colhead{$(H-K_{s})_{0}$} & \colhead{$J-K_{s}$} & \colhead{$H-K_{s}$} & \colhead{$A^{J-K_{s}}_{K_{s}}$} & \colhead{$A^{H-K_{s}}_{K_{s}}$} & \colhead{$\overline{A_{K_{s}}}$} & \colhead{$M_{K_{s}}$} & \colhead{BC$_{K_{s}}$} & \colhead{$Log~L_{bol}$}     \\
\colhead{(CXOGC J)} & \colhead{Type} & \colhead{(mag)} &\colhead{(mag)} & \colhead{(mag)} & \colhead{(mag)} & \colhead{(mag)} & \colhead{(mag)}& \colhead{(mag)}& \colhead{(mag)}& \colhead{(mag)}& \colhead{(mag)} &\colhead{($L_{\odot}$)} }
\startdata
 174502.8$-$290859 & O9I--B0I  & 9.88 & $-$0.21 & $-$0.10 &  4.05 &  1.55 &  2.10 &  2.38 &  2.24 & $-$6.86 &$-$3.7  &  6.12   \\
 174508.9$-$291218  &    WN7  &11.08 &  0.36   &    0.26 &  4.04 &  1.53 &  1.82 &  1.83 &  1.83 & $-$5.25 &$-$3.5    &5.40    \\
 174516.1$-$284909 & Ofpe/WN9  & 7.89 &  0.13   &    0.11 &  3.60 &  1.28 &  1.71 &  1.68 &  1.70 & $-$8.31 &$-$2.9   & 6.38   \\ 
 174516.7$-$285824 &  WN7--8h  &11.09 &  0.13   &    0.11 &  5.58 &  1.99 &  2.69 &  2.71 &  2.70 & $-$6.11 &$-$4.3    &6.06   \\
 174522.6$-$285844 & WN5--6b   & 12.22    & 0.36    & 0.26     & \nodata &2.99& \nodata    & 3.93   & 3.93   &   $-$6.21  & $-$3.8  & 5.90    \\
 174528.6$-$285605 &  WN8--9h  & 9.72 &  0.13   &    0.11 &  4.74 &  1.74 &  2.28 &  2.35 &  2.31 & $-$7.09 &$-$4.3   & 6.46     \\ 
 174532.7$-$285126  &  O4--6I  &10.72 & $-$0.21 & $-$0.10 &  3.95 &  1.41 &  2.06 &  2.17 &  2.11 & $-$5.89 &$-$4.4   & 6.02   \\  
 174536.1$-$285638 &  WN8--9h  &10.42 &  0.13   &    0.11 &  5.13 &  1.84 &  2.47 &  2.49 &  2.48 & $-$6.56 &$-$4.3   & 6.24 \\ 
174537.3$-$285354 &  O9--B0I  &11.23 & $-$0.21 & $-$0.10 &  4.52 &  1.57 &  2.34 &  2.40 &  2.37 & $-$5.64 &$-$3.7   & 5.64   \\  
174549.7$-$284925 & OB &            12.16 & $-$0.21 & $-$0.10 & \nodata & 2.10 &\nodata & 3.17 & 3.17& $-$5.51& $-$3.7 & 5.58\\ 
174550.2$-$284911 & WN8--9h & 10.47 &     0.13 & 0.11        &       4.77  &  1.69 & 2.29 & 2.28 & 2.29     & $-$6.32& $-$4.3  & 6.15\\ 
174550.4$-$284922 & WN8--9h &  9.87 &        0.13 & 0.11     &       5.07 & 1.77   & 2.44 & 2.39 &  2.42    & $-$7.05& $-$4.3  & 6.44 \\ 
174550.4$-$284919 & WN8--9h & 9.56 &         0.13 &  0.11     &      4.92 & 1.89   & 2.37 & 2.56  & 2.46    & $-$7.40& $-$4.3  &6.58\\ 
 174550.6$-$285617  &    WN7  &10.85 &  0.36   &    0.26 &  4.59 &  1.63 &  2.09 &  1.97 &  2.03 & $-$5.68 &$-$3.5   & 5.57  \\
 174555.3$-$285126   &   WN5--6b  &10.97 &  0.36   &    0.26 &  4.46 &  1.56 &  2.03 &  1.87 &  1.95 & $-$5.48 &$-$3.8   & 5.61    \\
174614.5$-$284937   & OBI               & 10.7  & $-$0.21 & $-$0.10 & 4.2 & 1.9 & 2.18 & 2.88     & 2.53 & $-$6.33  &$-$3.7 & 5.91 \\
174615.1$-$284932   &  OBI          & 10.59&$-$0.21 & $-$0.10  &    3.51   &  0.51  &   1.84 & 0.88 & 1.36   & $-$5.27 & $-$4.3  & 5.73\\ 
174616.6$-$284909   & O8--9I   & 11.08   & $-$0.21 & $-$0.10 & 5.45 & 1.91   &  2.80 & 2.89 & 2.85 &  $-$6.27 & $-$3.7 & 5.89 \\ 
 174617.0$-$285131  &   O6If$^+$  &10.49 & $-$0.21 & $-$0.10 &  4.49 &  1.63 &  2.32 &  2.49 &  2.41 & $-$6.42 &$-$4.3   & 6.19  \\ 
 174628.2$-$283920  &  O4--6I  &11.49 & $-$0.21 & $-$0.10 &  5.50 &  1.87 &  2.82 &  2.84 &  2.83 & $-$5.84 &$-$4.4  &  6.00 \\   
 174656.3$-$283232 &  WN8--9h  &11.24 &  0.13   &    0.11 &  0.00 &  2.50 &  0.00 &  3.44 &  3.44 & $-$6.70 &$-$4.3  &  6.30   \\ 
 174703.1$-$283119   &  O4--6I  &11.27 & $-$0.21 & $-$0.10 &  4.96 &  1.76 &  2.55 &  2.68 &  2.62 & $-$5.85 &$-$4.4  &  6.00  \\ 
 174711.4$-$283006 &  WN8--9h  &10.54 &  0.13   &    0.11 &  6.02 &  2.18 &  2.91 &  2.98 &  2.95 & $-$6.91 &$-$4.3   & 6.38    \\
 174712.2$-$283121 &  WN7--8h  &10.78 &  0.13   &    0.11 &  6.28 &  2.29 &  3.04 &  3.14 &  3.09 & $-$6.81 &$-$4.3   & 6.34   \\ 
 174713.0$-$282709  &  WN7--8h  &11.86 &  0.13   &    0.11 &  0.00 &  2.36 &  0.00 &  3.24 &  3.24 & $-$5.88 &$-$4.3   & 5.97   \\ 
 174725.3$-$282709   &  O4--6I  &11.30 & $-$0.21 & $-$0.10 &  0.00 &  2.07 &  0.00 &  3.12 &  3.12 & $-$6.32 &$-$4.4   & 6.19  \\ 
\enddata
\tablecomments{Intrinsic colors and BC$_{K}$ values were adopted from Crowther et al. (2006a) for all WN X-ray sources. For the O4--6I stars, intrinsic colors and BC$_{K}$ values were adopted from Martins \& Plez (2006) by averaging the values for spectral type O4--6I, and adopting the values for an O9.5I star to represent that of the O9I--B0I stars. We did not include WCd or DWCL stars in this analysis, since we do not have knowledge of their intrinsic stellar colors.}
\end{deluxetable}
\end{landscape}
\begin{deluxetable}{lccccccccc}
\setlength{\tabcolsep}{0.05in}
\renewcommand{\arraystretch}{0.80}
\tabletypesize{\scriptsize}
\tablecolumns{9}
\tablewidth{0pc}
\tablecaption{Free-free Model Parameters}
\tablehead{
\colhead{Star} & \colhead{Spectral} & \colhead{$T_{eff}$} & \colhead{$R^{*}$} & \colhead{$F_{\textrm{ff}}$ (3.6, 4.5, 5.8, 8.0 $\micron$)} &  \colhead{$T_{e}$} &  \colhead{$N_{e}\times10^{11}$} & \colhead{$R_{\textrm{ff}}$} & \colhead{$\lambda_{\tau=1}$}  \\
\colhead{} & \colhead{Type} & \colhead{(kK)} &  \colhead{($R_{\odot}$)} & \colhead{(Jy)} & \colhead{(K)} & \colhead{(cm$^{-3}$)} & \colhead{($R_{\odot}$)} &  \colhead{($\micron$)}   
} 
\startdata
X174516.1 & Ofpe/WN9 & 20 & 79 & 1.60~:~1.74~:~1.87~:~2.00 & 8100 & 0.8 &  246 & 5.2 \\
X174656.3 & WN8--9h & 30 & 33 & 0.29~:~0.26~:~0.21~:~0.14 & 7000 & 1.3 & 131  & 4.7  \\
 X174712.2 & WN7--8h & 35 & 24 & 0.31~:~0.31~:~0.26~:~0.18 &  8500 & 1.1 & 145  & 5.9 \\
 X174555.3 & WN5--6b & 40 & 12 & 0.16~:~0.16~:~0.14~:~0.10 & 11000 & 1.5 & 94  & 6.4 

\enddata
\end{deluxetable}

 \begin{landscape}
\begin{deluxetable}{lcccrcrcrrccccc}
\tablecolumns{15}
\setlength{\tabcolsep}{0.05in}
\renewcommand{\arraystretch}{1.4}
\tablewidth{0pc}
\tabletypesize{\tiny}
\tablecaption{\textit{Chandra} X-ray  Data for Massive Stellar X-ray Sources in the Galactic Center}
\tablehead{\colhead{X-Ray Source}& \colhead{R.~A.$_{\textrm{\tiny{X}}}$} & \colhead{Decl.$_{\textrm{\tiny{X}}}$} & \colhead{$\sigma_{\textrm{\tiny{X}}}$}&\colhead{Exposure} & \colhead{$C_{\textrm{\tiny{net}}}^{\textrm{\tiny{~soft}}}$} &  \colhead{$C_{\textrm{\tiny{net}}}^{\textrm{\tiny{~hard}}}$}&  \colhead{$F_{\textrm{\tiny{tot}}}$ (0.5--8 keV)}  &  \colhead{HR0} & \colhead{HR2} & \colhead{$\langle{E}\rangle$} & \colhead{Log$L_{\textrm{\tiny{X}}}$} & \colhead{Spectral} & \colhead{Ref.\tablenotemark{a}} & \colhead{Ref.\tablenotemark{b}} \\
\colhead{(CXOGC J)} & \multicolumn{2}{c}{(deg, J2000)}  &\colhead{(arcsec)} & \colhead{(sec)} & \colhead{} & \colhead{}  & \colhead{(cm$^{-2}$ s$^{-1}$)} & \colhead{} & \colhead{} & \colhead{(keV)} &  \colhead{(erg s$^{-1}$)} & \colhead{Type} & \colhead{X}& \colhead{IR} \\ [3pt]
\colhead{(1)} & \colhead{(2)} & \colhead{(3)} & \colhead{(4)} & \colhead{(5)} & \colhead{(6)} & \colhead{(7)} & \colhead{(8)} & \colhead{(9)} & \colhead{(10)} & \colhead{(11)} & \colhead{(12)} & \colhead{(13)} & \colhead{(14)} & \colhead{(15)} }
\startdata
174508.9$-$291218 & 266.28731 & $ -29.20474 $& 0.87 &  96809 &$<5.8$&$59.6_{-14.6}^{+15.2}$&  $3.39\times10^{-6}$&   $ 1.00_{-0.95}^{  -9.00} $ & $0.37_{-0.24}^{+0.24} $& 5.3 & 32.9 &  WN7 & 1 & 1 \\  

174516.1$-$290315 & 266.31750 & $ -29.05430 $& 0.44 & 936670  &$132.2_{-21.3}^{+22.8} $& $748.0_{42.5}^{42.5}$&   $3.42\times10^{-6} $&   $0.51_{-0.07}^{+0.07} $& $-0.33_{-0.10 }^ {+0.10} $& 3.3 & 32.0 & Ofpe/WN9 & 2 & 2\\ 

174516.7$-$285824 & 266.31959 & $ -28.97368 $& 0.65 &  1011840 & $<7.1$&$66.0_{-22.0}^{+19.8}$&     $3.56\times10^{-7}$&    $1.00_{  -0.59}^{  -9.00} $ &$-0.26_{  -0.44}^{   +0.40}$&  4.0 & 32.0 & WN7--8h &1 & 1 \\ 

174519.1$-$290321 & 266.32983 & $-29.05605 $ & 0.51 & 1011810 &$<14.3$ &  $197.5_{-33.6}^{+26.5}$&  $1.05\times10^{-6}$ & $1.00 _{-0.37}^{-9.00}$ &  $0.16_{-0.16}^{+0.17} $ & 5.0 & 32.4 & WC9d& 1 & 1 \\ 

174522.6$-$285844 & 266.34457 & $-28.97898$ & 0.63 & 1011840 & $<6.3$ &  $41.3_{-16.2}^{+11.4}$ & $2.30\times10^{-7}$ & $1.00 _{-0.95}^{-9.00}$ & $-0.33_{-0.44}^{+0.34}$& 4.4 & 31.8 & WN5--6b & 1 & 1 \\ %

174528.6$-$285605 & 266.36926 & $ -28.93483 $& 0.48 & 1011840  &$8.3_{-5.6}^{+6.9}$ &  $178.3_{-25.8}^{+21.0}$&   $9.00\times10^{-7}  $&  $0.82_{-0.13}^{+0.12} $ &$-0.32_{-0.18 }^ {+0.18} $& 3.7 & 32.4 &  Of& 2 & 6, 8 \\ 

174532.7$-$285617 & 266.38650 & $ -28.93807 $& 0.50 &  1011840 & $8.5_{-7.4}^{+6.0}$& $158.5_{-25.6}^{+22.4}$&   $7.18\times10^{-7}  $&  $0.75_{   -0.16}^{   +0.22}  $&$-0.13_{   -0.19 }^ { +0.21}$ & 3.9& 32.3 & O4--6I & 1 & 1 \\  

174536.1$-$285638 & 266.40059 & $ -28.94407 $& 0.37 & 1011830 &$216.7_{-27.6}^{+22.0}$  &$ 4860.0_{-91.0}^{+91.0}$&    $1.97\times10^{-5}$ &  $ 0.77_{   -0.02}^{   +0.03}$  &$-0.20_{   -0.02 }^ { +0.02} $& 3.9 & 33.7 & WN8--9h&  3 & 3 \\

174537.3$-$285354 & 266.40558 & $ -28.89844 $& 0.74 &  1011830 &$12.5_{-7.6}^{+10.5}$   & $91.0_{-22.9}^{+24.6}$&    $4.17\times10^{-7}$ &   $0.51_{   -0.31}^{   +0.29} $ &$-1.00_{   -9.00 }^{  +0.64} $& 3.4 & 32.0& O9I--B0I& 1 & 1 \\ 

174549.7$-$284925 & 266.45722&  $-28.82385 $& 0.41 & 110174 &  $<3.6$ & $42.1_{-11.4}^{+13.0}$& $1.62\times10^{-6}$ & $1.00_{-0.66}^{-9.00}$& $0.11_{-0.32}^{+0.31}$& 4.9 & 32.6 & OB & 4 & 6 \\  

174550.2$-$284911 & 266.45939 & $-28.81994 $ & 0.33 & 110174 &$25.7_{-8.1}^{+8.5}$  & $649.1_{-32.9}^{+32.9}$ & $3.00\times10^{-5}$& $0.77_{-0.07}^{+0.07}$&$-0.02_{-0.08}^{+0.08}$& 4.1 & 33.9 & WN8--9h& 4& 6, 8, 9 \\

174550.4$-$284922 & 266.46004 & $ -28.82285 $& 0.33 &   110174 &$43.4_{-10.3}^{+11.4}$   & $958.0_{-40.0}^{+40.0}$&     $3.97\times10^{-5}$&   $ 0.77_{   -0.06}^{   +0.05} $ &$-0.11_{    -0.07}^{   +0.07} $ & 4.0 & 34.0 & WN8--9h & 4 & 6, 8, 9\\ 
 
 174550.4$-$284919 & 266.46025 & $-28.82210 $ & 0.33 & 110174 &$32.5_{-8.9}^{+9.8}$   & $581.4_{-31.3}^{+31.3}$&  $2.29\times10^{-5}$& $0.76_{-0.06}^{+0.06}$&$-0.27_{-0.09}^{+0.09}$& 3.7 & 33.8 & WN8--9h& 4 & 6, 8, 9 \\  

174550.6$-$285919 & 266.46088 & $ -28.98879 $& 0.43 & 1011850 &$11.5_{-8.1}^{+8.9}$ & $43.7_{-21.7}^{+16.0}$&    $2.06\times10^{-7} $& $ -1.00_{  -9.00}^{  +1.21}  $&$-0.14_{  -0.58 }^ { +0.46} $& 4.0 & 31.7 & WN7 & 2 & 8\\ 

174555.3$-$285126 & 266.48075 & $ -28.85738 $& 0.49 &  208788 &$<5.6$   & $86.2_{-17.3}^{+18.6}$&  $1.65\times10^{-6} $&  $ 1.00_{   -0.30}^{  -9.00} $& $-0.64_{  -0.28 }^ { +0.25} $ & 3.5 & 32.6 & WN5--6b& 1 & 1 \\ 

174614.5$-$284937 & 266.56058 & $ -28.82698 $& 0.50 &  158922  &$46.1_{-10.3}^{+12.0}$   & $20.7_{-8.4}^{+8.8}$&    $1.86\times10^{-6} $&  $-0.60_{   -0.18}^{   +0.18} $ &$-1.00_{  -9.00}^{   +0.90} $ &2.1 & 32.7 & OBI & 4 & 11 \\ 

174614.6$-$284940 & 266.56113 & $ -28.82794 $& 0.62 &  158922  &$<3.4$  & $27.9_{-9.7}^{+9.6} $&   $ 8.90\times10^{-7}$ &   $1.00_{   -0.32}^{  -9.00} $ &$-1.00_{  -9.00}^{   +0.82} $ &3.3 & 32.3 & DWCL& 4& 7 \\ 

174615.1$-$284932 & 266.56311 & $ -28.82578 $& 0.51 & 158922  &$<5.9$   & $61.4_{-15.1}^{+16.1}$&    $1.36\times10^{-6} $&   $1.00_{   -0.23}^{  -9.00}  $&$-0.51_{   -0.36 }^{  +0.35} $ &3.7 & 32.5 & OBI& 4 & 7\\ 

174615.8$-$284945 & 266.56610 & $ -28.82931 $& 0.49 &  158922 &$<4.5$  & $69.9_{-15.8}^{+15.8}$&    $1.80\times10^{-6}$ &   $1.00_{   -0.73}^{  -9.00} $ &$-0.22_{   -0.23 }^{  +0.23} $ &4.5 & 32.7 & DWCL& 4 & 7 \\ 

174616.6$-$284909 & 266.56939 & $ -28.81920 $& 0.75 &  208347 &$<6.0$   & $31.1_{-16.0}^{+17.7}$  &  $4.91\times10^{-7} $&   $1.00_{   -0.48}^{  -9.00} $ &$-9.00_{  -9.00}^{  -9.00} $& 2.9 & 32.1 & O8--9I& 4 & 7 \\ 

174617.0$-$285131 & 266.57102 & $ -28.85867 $& 0.56 & 208347 &$<8.4$  & $114.5_{-22.7}^{+22.5}$&    $2.38\times10^{-6}$ &  $ 1.00_{   -0.22}^{  -9.00} $& $-0.27_{   -0.26 }^{  +0.25} $& 3.9 & 32.8& O6If$^+$ & 1 & 1 \\ 

174617.7$-$285007 & 266.57417 & $ -28.83537 $& 0.50 &  158922 &$<4.6$   & $70.2_{-15.6}^{+17.0}$&    $1.60\times10^{-6}$ &   $1.00_{   -0.45}^{  -9.00}$  &$-0.48_{   -0.27}^ {  +0.28} $ &3.5 & 32.6 & WC9d& 1 & 1 \\ 
 
174628.2$-$283920 & 266.61791 & $ -28.65569 $& 0.51 &  49481 &$<2.5$ &  $24.0_{-8.2}^{+8.1}$&  $3.62\times10^{-6} $ &  $1.00_{   -0.52}^{  -9.00}  $&$ 0.12_{   -0.41 }^ { +0.41} $ &4.4 & 33.0 & O4--6I& 1 & 1 \\  

174645.2$-$281547 & 266.68842 & $ -28.26319 $& 1.04 &   31254 &$<3.5$  & $265.1_{-26.3}^{+27.9}$&  $ 6.83\times10^{-5} $&   $1.00_{   -0.38}^{  -9.00}$  &$ 0.25_{   -0.10  }^{ +0.10}  $&5.2 & 34.2 & DWCL& 5 &  5\\ 

174656.3$-$283232 & 266.73465 & $ -28.54230 $& 0.67 &  157384 &$<4.3$  & $135.8_{-20.8}^{+20.9}$&    $3.91\times10^{-6}$&   $ 1.00_{   -0.41}^{  -9.00}$ &$ -0.09_{  -0.17}^ {  +0.17}$ & 4.4 & 33.0 & WN8--9h& 1 & 1\\ 

174703.1$-$283119 & 266.76298 & $ -28.52222 $& 0.7 &   107903 & $<2.4 $ & $42.4_{-10.7}^{+11.7}$&    $1.83\times10^{-6}$ &   $1.00_{ -0.45}^{  -9.00}$  & $-0.16_{  -0.30 }^  {+0.31}  $&4.2 & 32.7 & O4--6I& 1 & 1 \\ 

174711.4$-$283006 & 266.79763 & $ -28.50194 $& 0.60 &  118333 &$<2.37$ & $36.12_{-9.8}^{+11.3}$&     $1.13\times10^{-6}$&   $ 1.00_{   -0.39}^{  -9.00}$ &$ -0.55_{  -0.32}^ {  +0.35} $ &3.6 & 32.4 &  WN8--9h& 1 & 1 \\ 

174712.2$-$283121 & 266.80097 & $ -28.52258 $& 0.71  &  118333 &$<2.5$   & $45.3_{-11.4}^{+12.6}$&   $1.74\times10^{-6}$ &  $ 1.00_{   -0.66}^{  -9.00}$ & $-0.19_{  -0.32 }^ { +0.29}  $&4.5 & 32.6 & WN7--8h& 1 & 1 \\ 

174713.0$-$282709 & 266.80425 & $ -28.45251 $& 0.45 &  107907 &$<1.7$   & $17.7_{-7.0}^{+7.4}$&   $6.77\times10^{-7} $ &  $1.00_{   -1.19}^{  -9.00}  $&$-0.27_{  -0.48  }^ {+0.39}$  &4.3 & 32.2 & WN7--8h& 1 & 1 \\ 

174725.3$-$282523 & 266.85559 & $ -28.42301 $& 0.60 &  128753 &$<3.5$   & $59.3_{-13.4}^{+14.5}$&   $ 2.21\times10^{-6} $&   $1.00_{   -0.92}^{  -9.00}$ &$ -0.02_{  -0.24 }^ { +0.25} $ &4.6 & 32.7 & O4--6I& 1 & 1 \\ 

\enddata
\tablecomments{The table columns contain the following: \\
(1) The source name, which was derived from the coordinates of the sources based on the IAU format, in which least-significant figures are \textit{truncated} (as opposed to rounded). The names should not be used as the locations of the sources. \\
(2--3) The right ascension and declination of the X-ray source, in degrees (J2000). \\
(4) The 95\%-confidence uncertainty in the position (the error circle). \\ 
(5) The total exposure time in seconds. \\
(6) The net number of counts in the soft 0.5--2.0 keV energy band, with 90\%-confidence uncertainties.   \\
(7) The net numbers of counts in the hard 2.0--8.0 keV energy band, with 90\%-confidence uncertainties.\\
(8) The total 0.5--8.0 keV flux in units of photons cm$^{-2}$ s$^{-1}$\\
(9) The soft color, defined as $\textrm{HR0}=(h-s)/(h+s)$, where $h$ and $s$ is the flux in the 2.0--3.3 keV and  0.5--2.0 keV bands, respectively. The upper and lower 90\%-confidence uncertainty is included. An HR  value of $-$9.00 indicates that the sources was not detected in either of the $h$ or $s$ bands. HR uncertainties of $-$9.00 are considered null, and are present whenever a source is detected in only one band.  \\
(10) The hard color, defined as $\textrm{HR2}=(h^{\prime}-s^{\prime})/(h^{\prime}+s^{\prime})$, where $h^{\prime}$ and $s^{\prime}$ is the flux in the 4.7--8.0 keV and  3.3--4.7 keV bands, respectively. The upper and lower 90\%-confidence uncertainty is included.\\
(11) The average energy per photon in units of keV. This value can be multiplied by the photon flux in column 8 for an estimate of energy flux. \\
(12) The estimated X-ray luminosity of the source (0.5--8.0 keV), assuming $N_{\textrm{\tiny{H}}}=6\times10^{22}$ cm$^{-2}$ and $kT=2$ keV. Owing to our insensitivity to soft X-ray photons between 0.5--2.0 keV, the uncertainties are significant, so, these values should be met with caution (see text).  
(13) The spectral type of the stellar counterpart. \\
(14) References to the first published association of a massive star with an X-ray source.  \\
(15) References to the spectral classification of a source. }
\tablenotetext{a}{The following authors first recognized a massive-star as an X-ray source: (1) This work or Mauerhan et al. 2007; (2) Muno et al. 2006b; (3) Mikles et al. 2006; (4) Yusef-Zadeh et al. 2002; (5) Hyodo et al. 2008.}
\tablenotetext{b}{The following authors determined the spectral type of the associated massive star from infrared observations: in addition to the authors listed in a), (6) Figer et al. 2002; (7) Figer et al. 1999; (8) Cotera et al. 1996 or Cotera et al. 1999; (9) Martins et al. 2008; (10) Clark et al. 2009; (11) Liermann et al.  2009.}
\end{deluxetable}
\end{landscape}

\begin{deluxetable}{lcccccccc}
\setlength{\tabcolsep}{0.06in}
\renewcommand{\arraystretch}{1.4}
\tabletypesize{\scriptsize}
\tablecolumns{9} 
\tablewidth{0pc}
\tablecaption{$L_{\textrm{\scriptsize{X}}}$ Compared with Other Authors}
\tablehead{ \colhead{Source}&  \colhead{$kT$ (us)} &  \colhead{$N_{\textrm{\tiny{H}}}$ (us)} & \colhead{Log$L_{\textrm{\tiny{X}}}$ (us)}  &  \colhead{$kT$ (them)} &  \colhead{$N_{\textrm{\tiny{H}}}$ (them)} & \colhead{Log$L_{\textrm{\tiny{X}}}$  (them)} & \colhead{$\Delta L_{\textrm{\tiny{X}}}$} & \colhead{Ref.}  \\
 \colhead{(CXOGC J)}  & \colhead{(keV)} & \colhead{(cm$^{-2}$)} & \colhead{(erg s$^{-1}$)} & \colhead{(keV)} & \colhead{(cm$^{-2}$)} & \colhead{(erg s$^{-1}$)} & \colhead{(dex)} & \colhead{}}
\startdata
174528.6$-$285605  &2 &6 & 32.4 & $1.2^{+0.2}_{-0.2}$ & $7.7^{+1.5}_{-1.5}$ & 33.1 & 0.7 & 2 \\
174550.2$-$284911 & 2&6 & 33.9 & $3.25^{+2.62}_{-1.24}$&$6.4^{+2.5}_{-1.6}$& 33.7 & 0.2 & 3 \\
174550.4$-$284922 & 2&6 &  34.0 &$ 2.1^{+0.58}_{-0.34}$& $8.1^{+1.1}_{-1.2}$ & 33.9 & 0.1 & 3 \\
174550.4$-$284919 & 2&6 & 33.8 &$1.87^{+0.39}_{-0.32}$ & $7.3^{+1.5}_{-1.1}$& 34.0 & 0.2 & 3 \\
174550.6$-$285919 & 2&6 & 31.7 & 2 &  6 & 31.7 & 0.0 & 2 \\
174645.2$-$281547 & 2&6 & 34.2 &$3.8^{+0.5}_{-0.6}$ &$23^{+3}_{-2}$& 34.5\tablenotemark{a}  & 0.3 & 1 
\enddata
\tablecomments{This table compares the X-ray luminosities derived via (1) X-ray photometry, adopting constant average values for plasma temperature ($kT$) and interstellar absorption ($N_{\textrm{\tiny{H}}}$), and (2) X-ray spectral fitting with temperature and absorption as free parameters. Only non-variable X-ray sources were used for this comparison. References: (1) Hyodo et al. 2008; (2) Muno et al. 2006; (3) Wang et al. 2006.} 
\tablenotetext{a}{The luminosity of this source is for the 2.0--8.0 keV band.}
\end{deluxetable}

\begin{landscape}
\begin{deluxetable}{lcccccccc}
\setlength{\tabcolsep}{0.06in}
\renewcommand{\arraystretch}{1.2}
\tabletypesize{\scriptsize}
\tablecolumns{9} 
\tablewidth{0pc}
\tablecaption{Massive Stars Exhibiting X-Ray Variability}
\tablehead{\colhead{Record} & \colhead{Source}&  \colhead{Min. ObsID} &  \colhead{UT Date} & \colhead{$F_{\textrm{\tiny{min}}}$} & \colhead{Max. ObsID} & \colhead{UT Date} & \colhead{$F_{\textrm{\tiny{max}}}$} & \colhead{$F_{\textrm{\tiny{max}}}/F_{\textrm{\tiny{min}}}$}   \\
\colhead{No.} & \colhead{(CXOGC J)}  & \colhead{} & \colhead{} & \colhead{($10^{-6}$ cm$^{-2}$ s$^{-1}$)} & \colhead{}& \colhead{}& \colhead{($10^{-6}$ cm$^{-2}$ s$^{-1}$)}& \colhead{}}
\startdata
 247 & 174516.1$-$290315 & 6113 & 2005 Feb 27 06:26:24 & $<$2.8        & 4684 & 2004 Jul 06 22:29:57  & $9.4\pm1.0$&     $3.0^{-1.0}_{-1.0}$  \\ 
1733 & 174536.1$-$285638  & 2952 & 2002 Mar 23 12:25:04 & $7.1\pm1.9$ & 5950 & 2005 Jul 24 19:58:27  & $24.0\pm1.5$&    $3.4^{1.3}_{0.7}$  \\ 
8598 & 174712.2$-$283121  &  944 & 2000 Mar 29 09:44:36 & $1.1\pm0.3$ & 2277 & 2001 Jul 16 11:52:55 & $4.1\pm1.9$&     $3.6^{2.1}_{1.8}$  \\ 
 7624 & 174614.5$-$284937  &  945 & 2000 Jul 07 19:05:19 & $1.7\pm0.9$ & 4500 & 2004 Jun 09 08:50:32 & $2.7\pm0.4$&    $1.6^{1.8}_{0.6}$    
\enddata
\tablecomments{The columns give: the record number from the catalog of Muno et al. (2009), the source name, the ObsID and UT date in which the minimum flux was observed, the minimum flux, the ObsID and UT date in which the maximum flux was observed, and the maximum flux. The final column contains the ratios of the maximum and minimum observed fluxes, with upper and lower uncertainties. Uncertainties of $-1.0$ imply that the flux ratio is a lower limit.} 
\end{deluxetable}
\end{landscape}

\begin{figure*}[t]
\centering
\epsscale{1}
\plotone{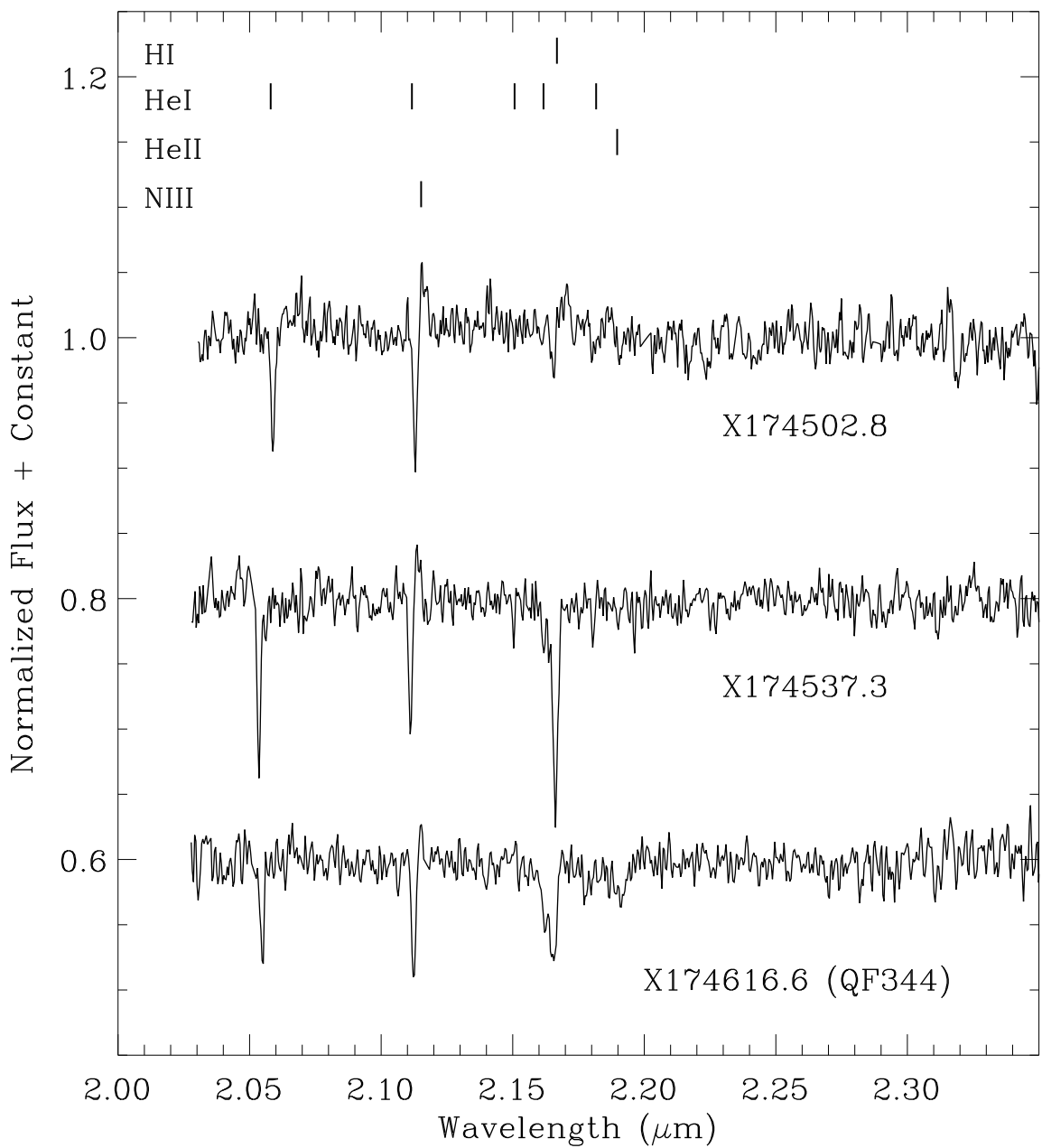}
\caption[$K$-band Spectra of Late O Supergiant X-ray Sources]{\linespread{1}\normalsize{$K$-band spectra of late-O/early-B supergiant X-ray sources in the Galactic center, exhibiting absorption lines of H {\sc i} and He {\sc i}, and a weak emission feature from N {\sc iii} near $\lambda$2.11 {\micron}. The spectra of X174537.3 and X174502.8 resemble that of O9I--B0I stars, while the presence of  He {\sc ii} absorption and a relatively shallow Br$\gamma$ line in the spectrum of X174616.6 are indicative of a slightly earlier spectral type in the range of O8--9I.}}
\end{figure*}

\begin{figure*}[t]
\centering
\epsscale{1}
\plotone{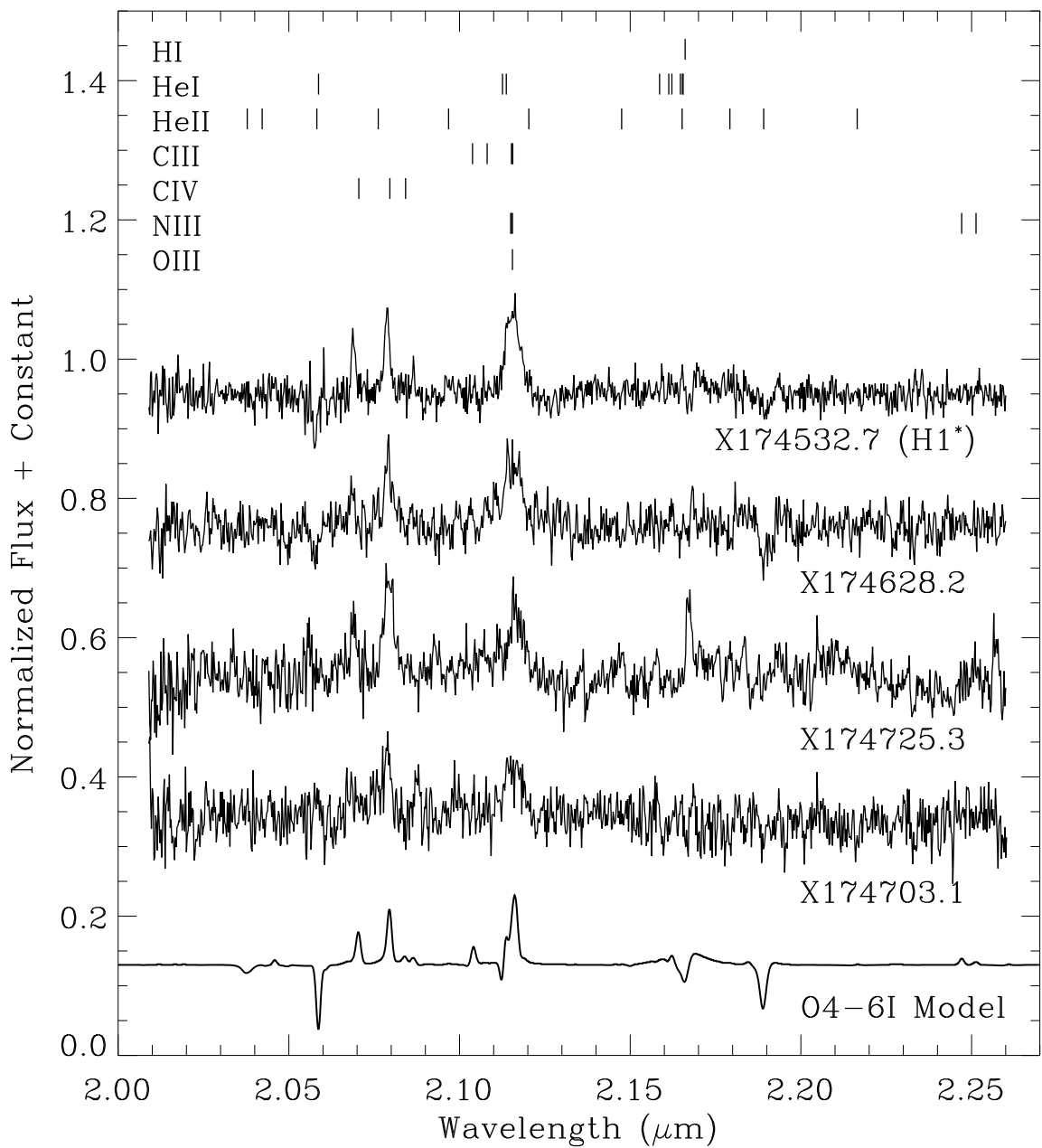}
\caption[$K$-band Spectra of Early Of Supergiant X-ray Sources]{\linespread{1}\normalsize{$K$-band spectra of early O4--6{\sc I} supergiant X-ray sources, dominated by emission from the complex of He {\sc i}, N {\sc iii}, C {\sc iii}, and O {\sc iii} near $\lambda$2.112--2.115 {\micron}. Weak features of Br$\gamma$ emission or absorption and He {\sc ii} absorption at $\lambda$2.19 {\micron} are also present. A model spectrum of an O4--6I star is plotted near the bottom of the figure for comparison (Martins et al. 2008). }}
\end{figure*}

\begin{figure*}[t]
\centering
\epsscale{1}
\plotone{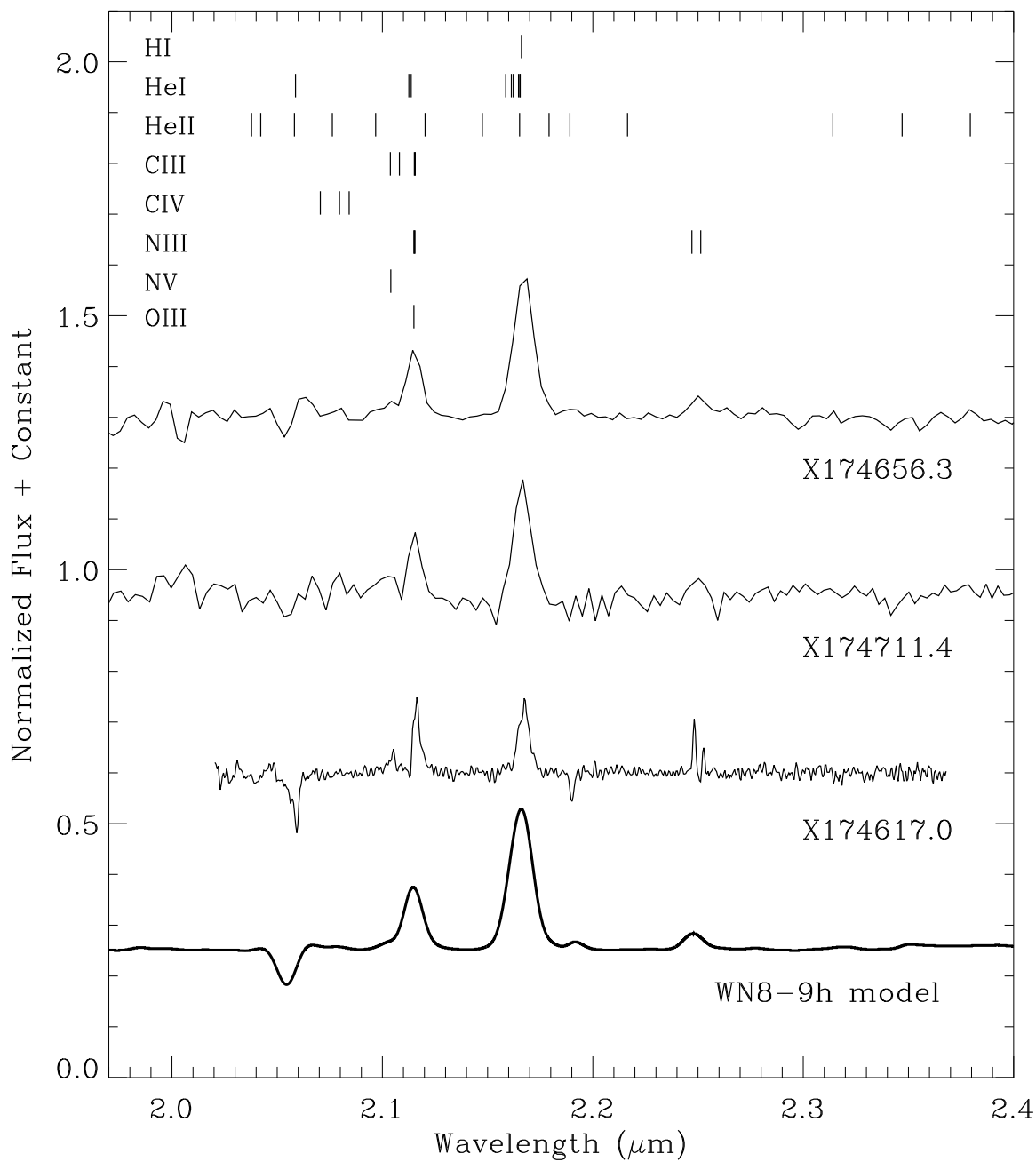}
\caption[$K$-band Spectra of WN8--9h X-ray Sources]{\linespread{1}\normalsize{$K$-band spectra of  WN8--9h X-ray sources X174656.3 and X174711.4, and the O6If$^+$ star X174617.0 from Mauerhan, Muno, \& Morris (2007). Strong Br$\gamma$ emission is accompanied by the emission complex of He {\sc i}, N {\sc iii}, C {\sc iii} and O {\sc iii} near $\lambda$2.112--2.115 {\micron}. A model spectrum of a WN8--9h star is plotted near the bottom of the figure for comparison (Martins et al. 2008). }}
\end{figure*}

\begin{figure*}[t]
\centering
\epsscale{1}
\plotone{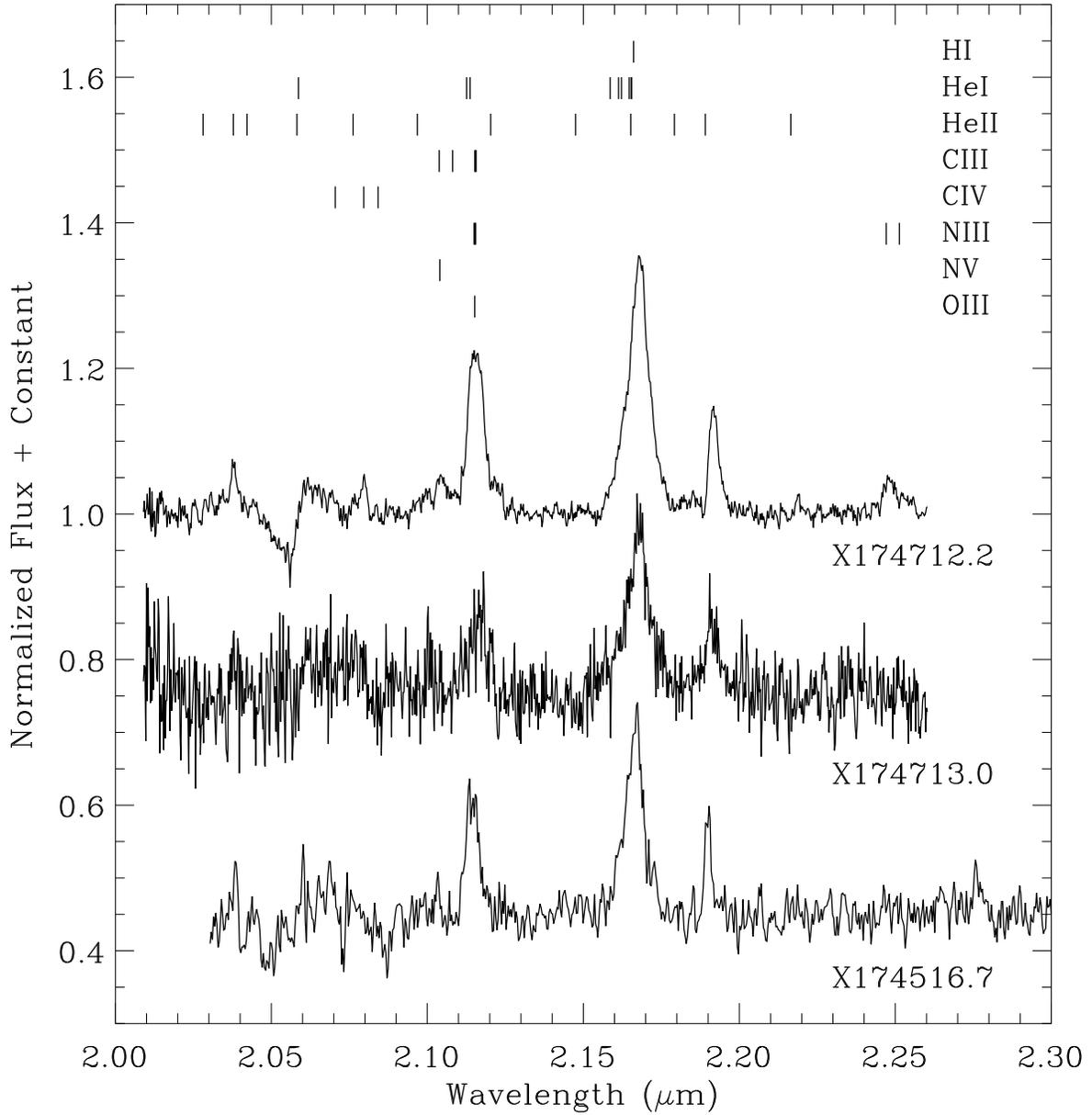}
\caption[$K$-band Spectra of WN7--8h X-ray Sources]{\linespread{1}\normalsize{$K$-band Spectra of WN7--8h X-ray sources, which exhibit stronger He {\sc ii} emission at $\lambda2.189$ {\micron}, relative to the WN8--9h stars presented in Figure 3. }}
\end{figure*}

\begin{figure*}[t]
\centering
\epsscale{0.9}
\plotone{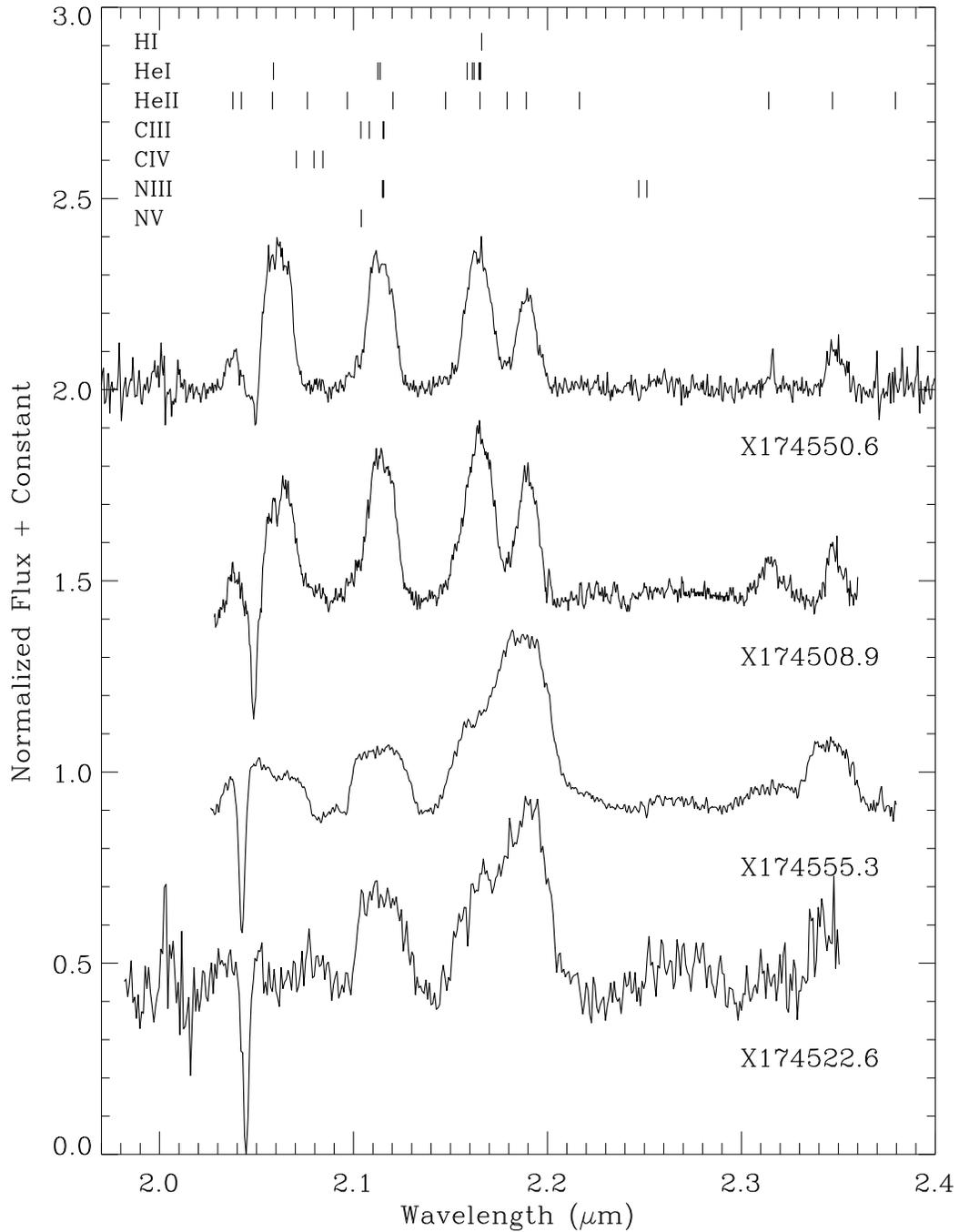}
\caption[$K$-band Spectra of WN X-ray Sources]{\linespread{1}\normalsize{$K$-band spectra of hydrogen-deficient Wolf-Rayet X-ray sources. Broad emission lines of He {\sc i} and He {\sc ii} dominate the spectra, indicating high temperatures and fast extended winds. X174550.6 and X174508.9 have spectral types of WN7; X174555.3 and X174522.6 have spectral types in the range of WN5--6b.}}
\end{figure*}

\begin{figure*}[t]
\centering
\epsscale{1}
\plotone{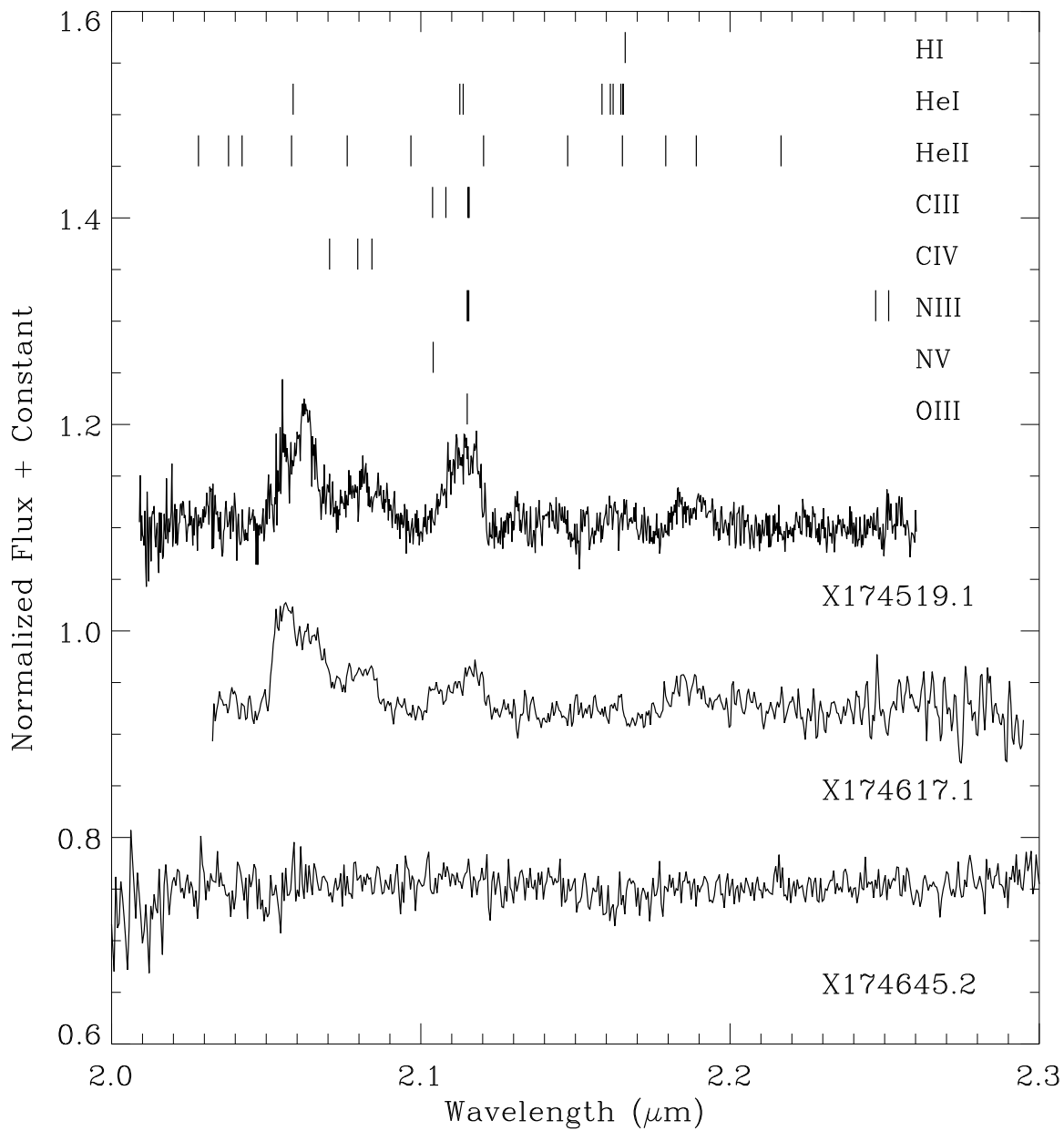}
\caption[$K$-band Spectra Dusty WC X-ray Sources]{\linespread{1}\normalsize{$K$-band spectra of WC counterparts to Galactic center X-ray sources. The weakness of the emission lines is attributable to dilution by strong excess continuum emission from hot dust, and perhaps from a supergiant companion. The spectrum of X174645.2 is an extreme example of dust dilution, and is completely featureless within the noise level.}}
\end{figure*}

\begin{figure*}[t]
\centering
\epsscale{1}
\plotone{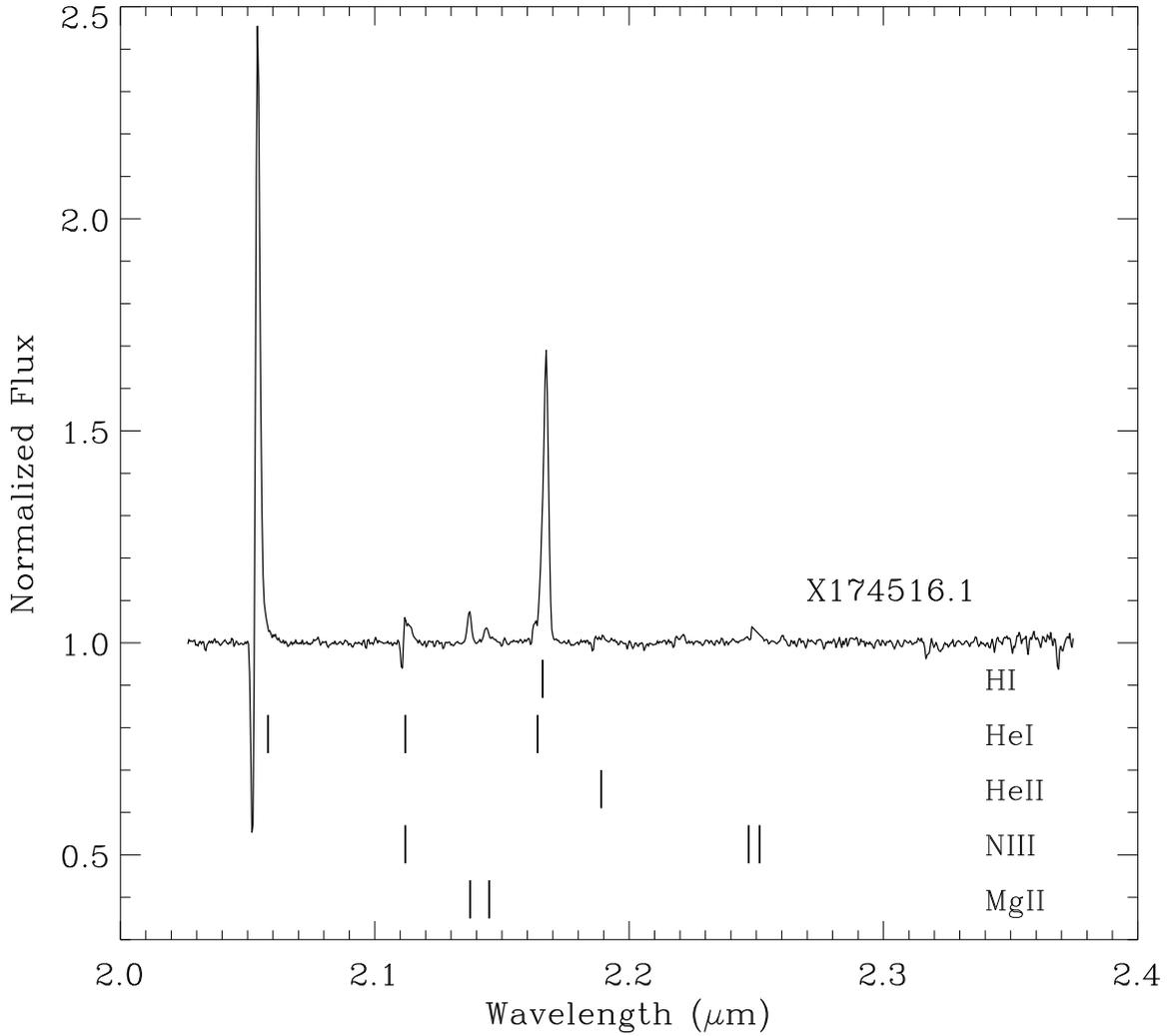}
\caption[$K$-band Spectrum of the CXOGC J174516.1$-$281531]{\linespread{1}\normalsize{$K$-band spectra of the counterpart to X-ray source CXOGC J174516.1$-$281531. Strong emission lines of Br$\gamma$ and He {\sc i} P-Cygni at $\lambda$2.058 {\micron}, are accompanied by weaker lines of Mg {\sc ii} near $\lambda$2.14 {\micron}, and weak He {\sc i} and He {\sc ii} P-Cygni profiles near $\lambda$2.11 {\micron} and $\lambda$2.19 {\micron}, respectively. The spectrum resembles that of Ofpe/WN9 stars.}}
\end{figure*}

\begin{figure*}[t]
\centering
\epsscale{1}
\plottwo{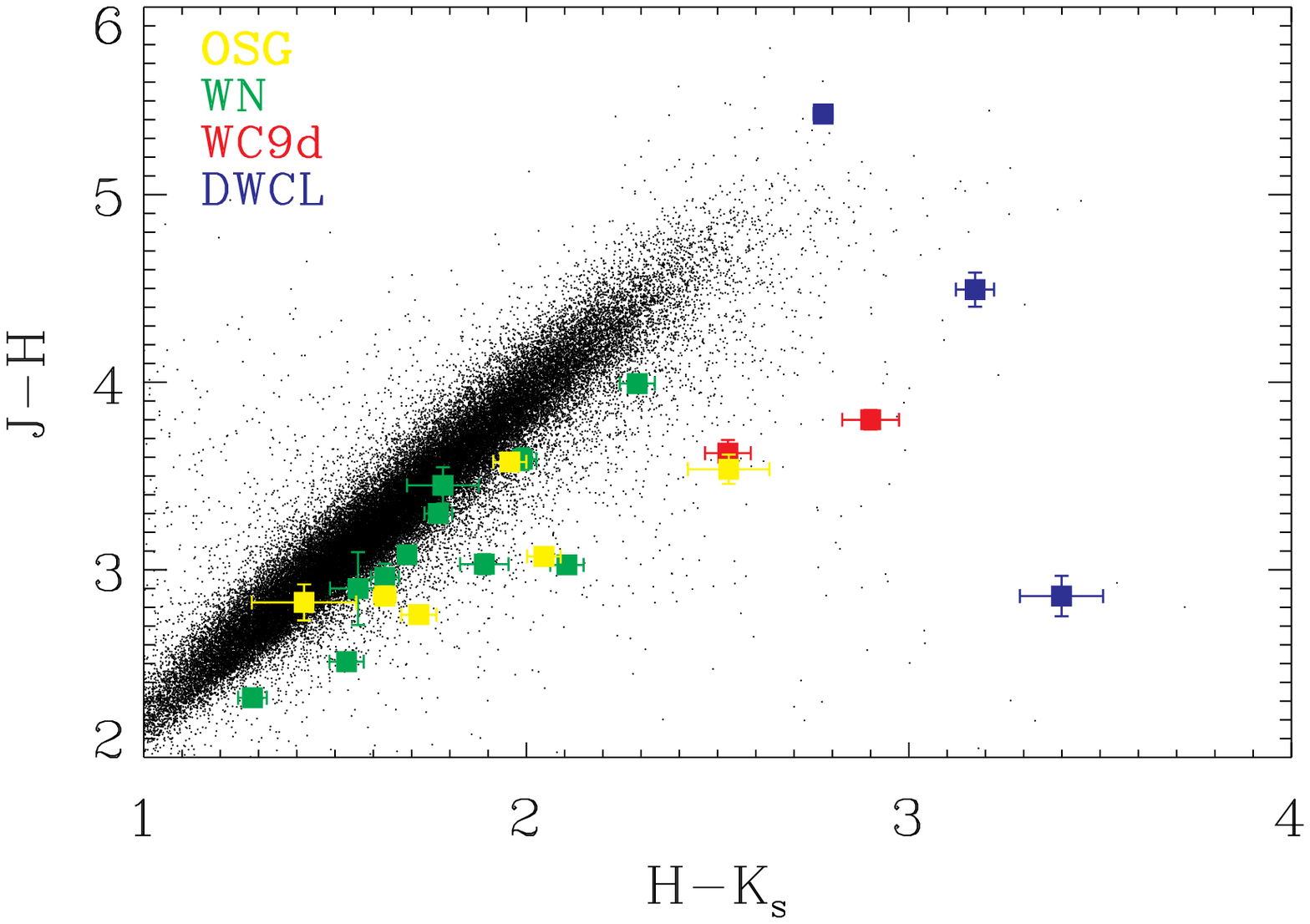}{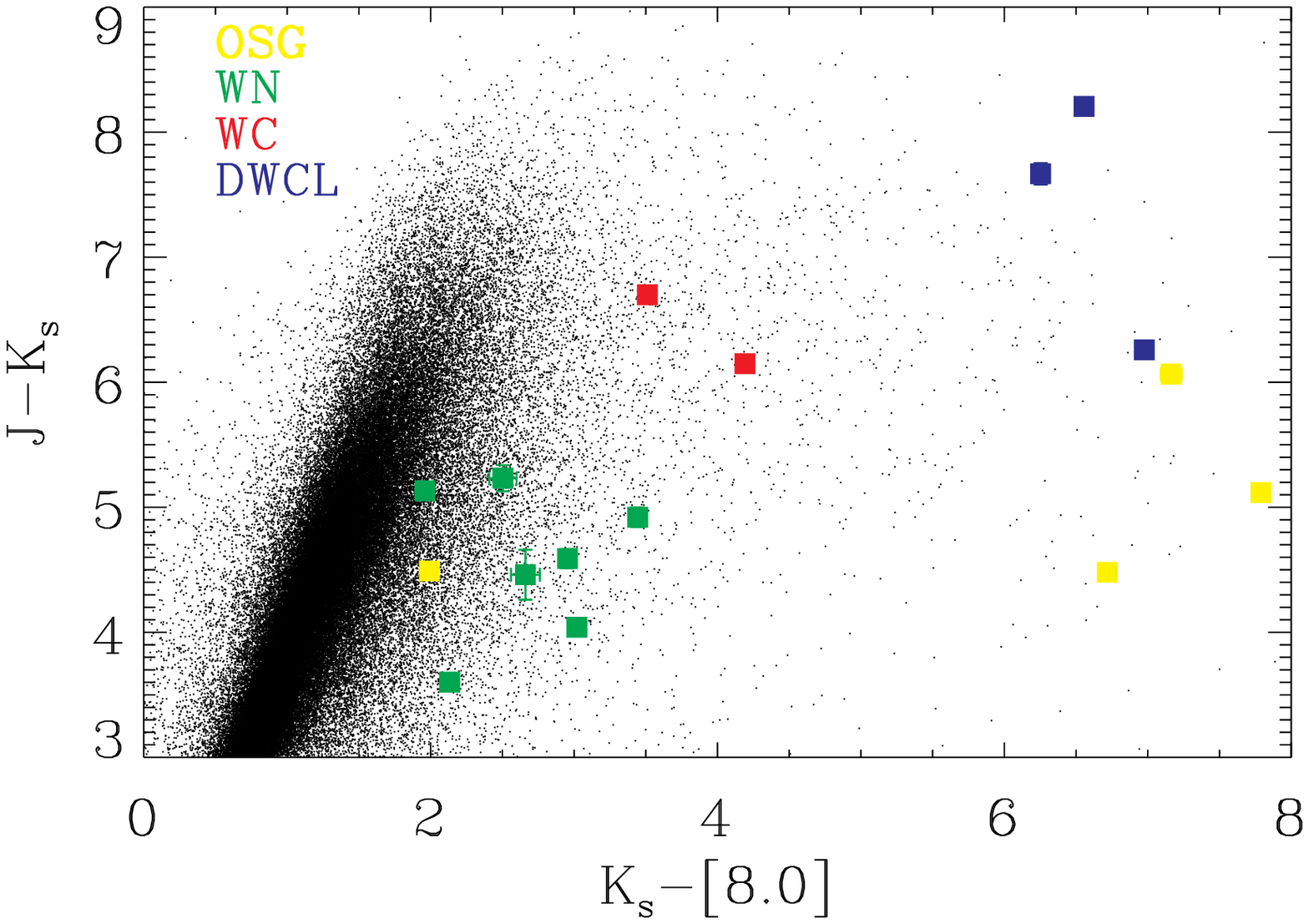}
\caption[Near-Infrared Color-color Diagram of Massive Stellar X-ray Sources]{\linespread{1}\normalsize{Infrared color-color diagram of massive stellar X-ray sources near the Galactic center, plotted with 1$\sigma$ error bars.  Almost all of the massive stars (large symbols) exhibit significant long-wavelength excess compared to field stars from the SIRIUS near-infrared survey (small dots).}}
\end{figure*}

\begin{figure*}[h]
\centering
\epsscale{1}
\plotone{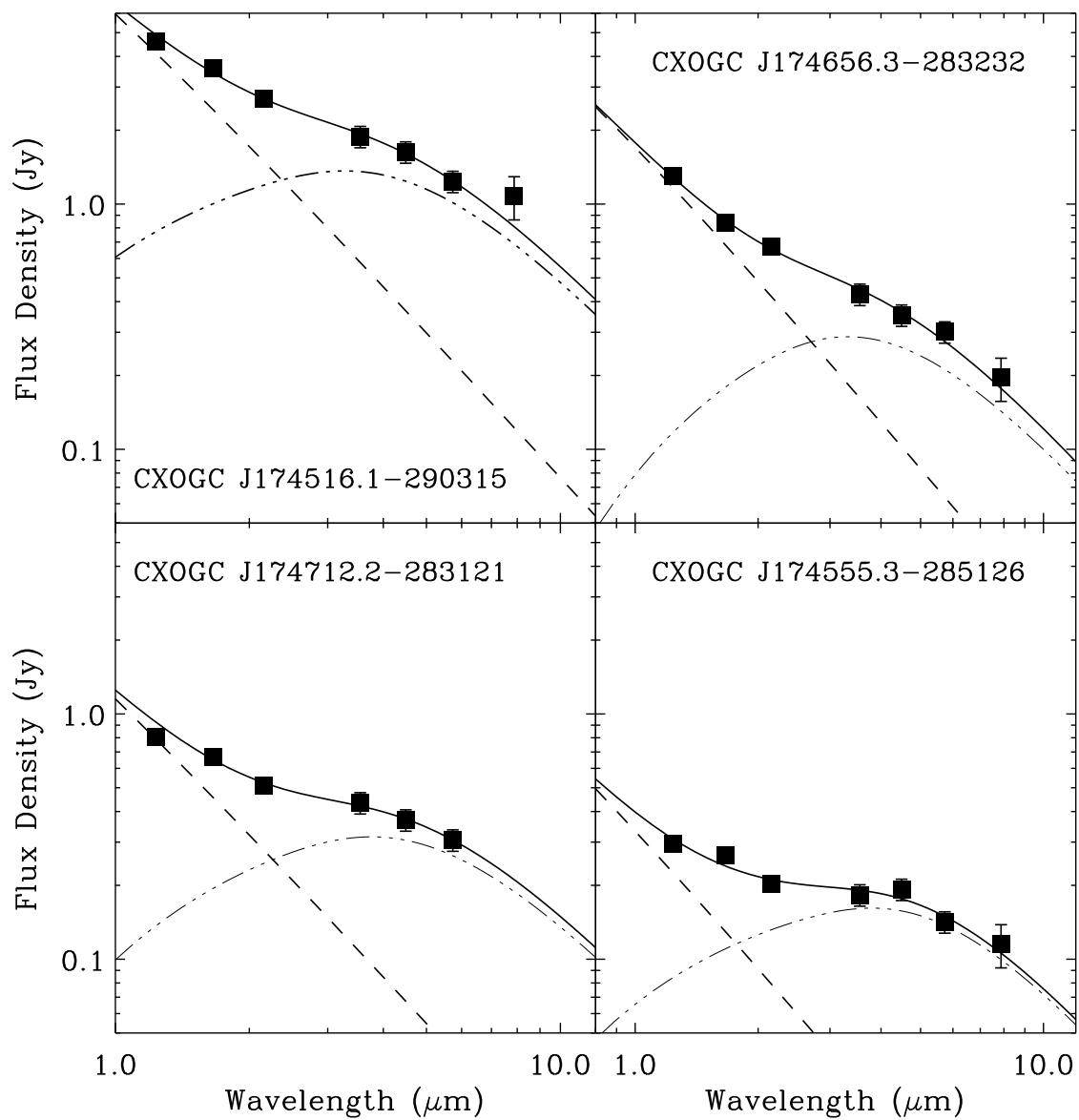} 
\caption[1--10 {\micron} SEDs of WN stars with Free-free Models.]{\linespread{1}\normalsize{Extinction-corrected $\lambda$1--10 {\micron} photometry (filled squares) and model SEDs for selected WN stars exhibiting free-free emission in the infrared. For the model SEDs, the dashed lines represent the Rayleigh--Jeans portion of the stellar continuum; the dash-dotted lines represent the free-free emission spectrum; and the solid lines are the model sum.}}
\label{archesquintuplet}
\end{figure*}

\begin{figure*}[t]
\centering
\epsscale{1}
\plotone{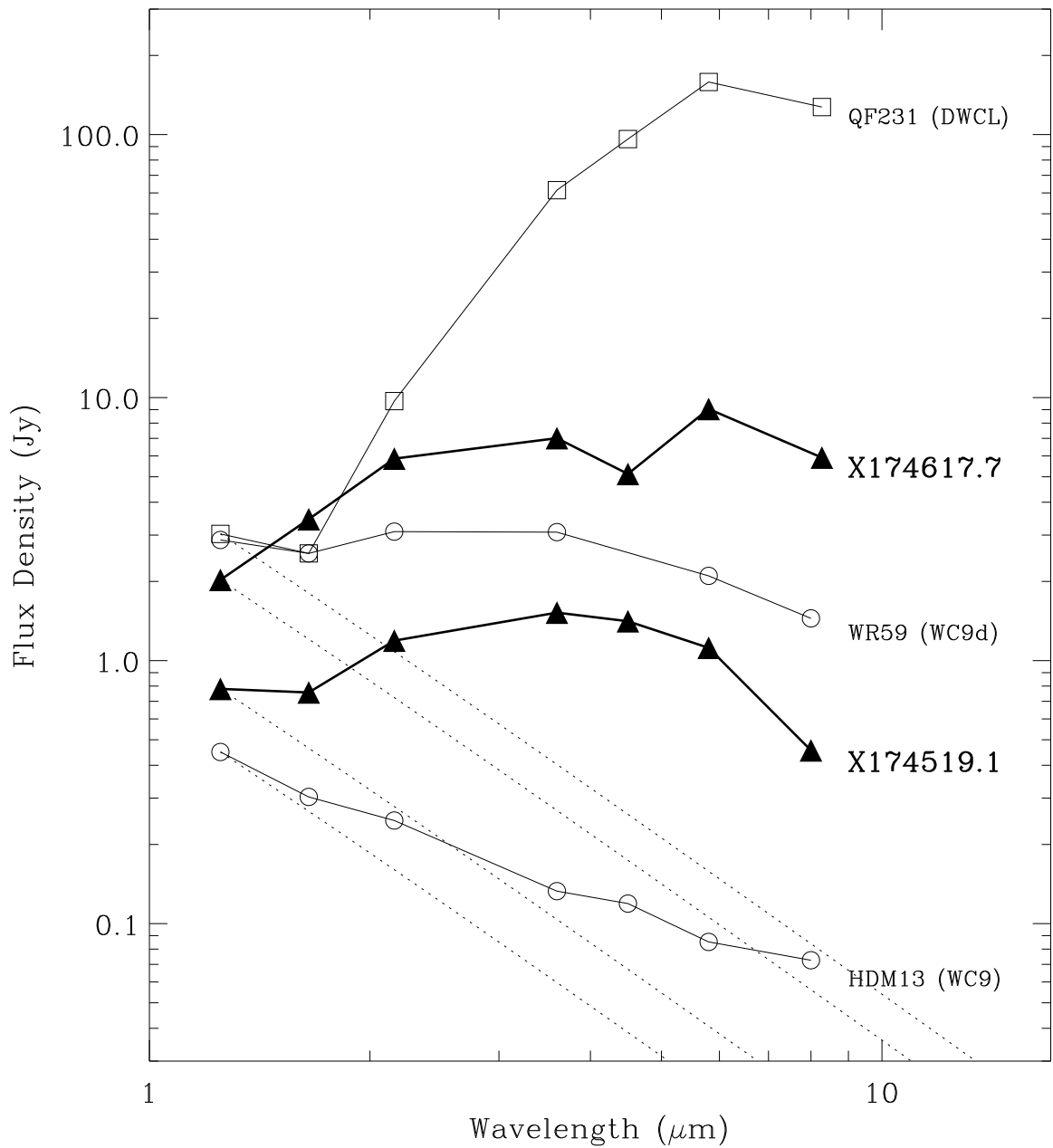}
\caption[$\lambda$1--10 {\micron} SEDs of dusty WC9 stars]{\linespread{1}\normalsize{Extinction-corrected $\lambda$1--10 {\micron} SEDs for new WC9d stars X174519.1 and X174617.7 (filled triangles), compared with the dust-enshrouded, extreme DWCL star qF231 of the Quintuplet cluster (squares), the single WC9 star HDM13 and the WC9d star WR 59 (circles). Rayleigh--Jeans curves (dotted lines) appropriate for WC9 stars ($T_{eff}=40$ kK) are scaled to the $J$-band data points for comparison with the SED curves. The SEDs of X174519.1 and X174617.7 exhibit an infrared excess similar to that of WR 59, consistent with the WC9d class.}}
\end{figure*}
\clearpage

\begin{figure*}[t]
\centering
\epsscale{1}
\plotone{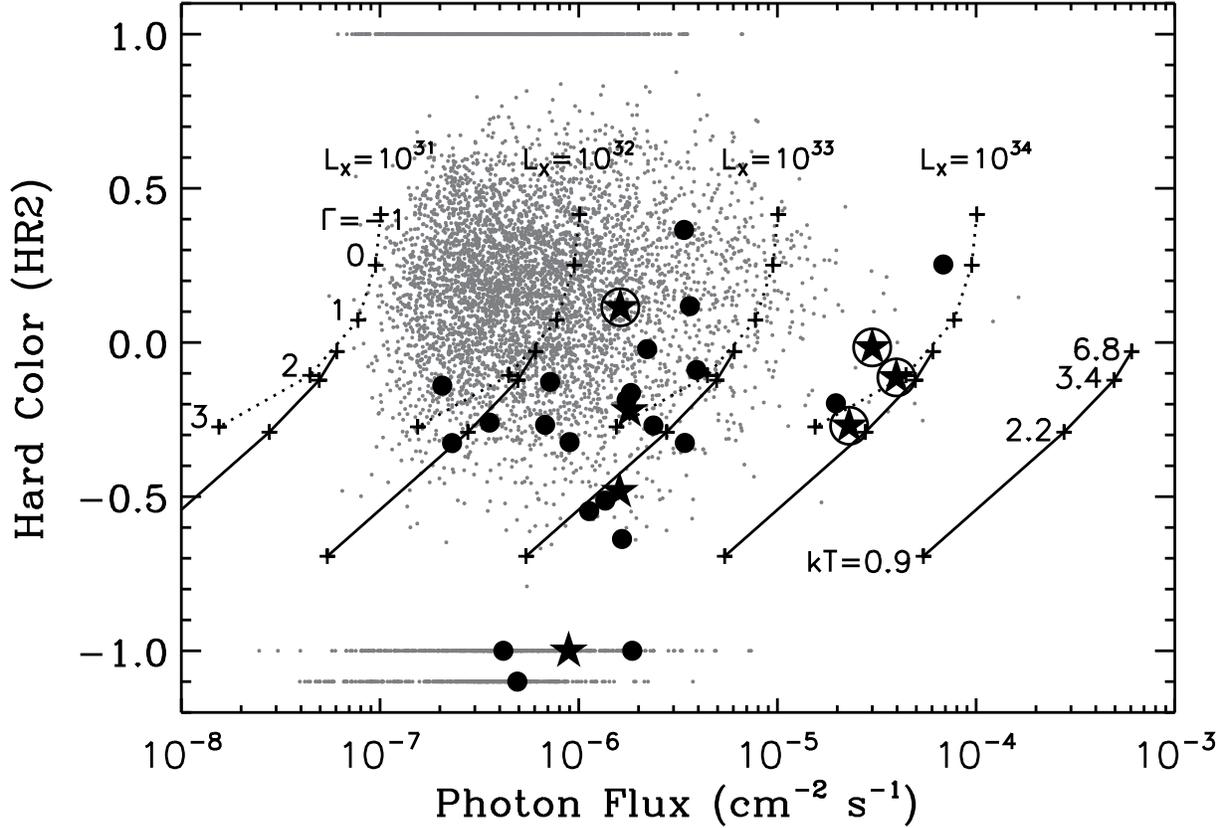}
\caption[X-Ray Flux vs. Color Diagram]{\linespread{1}\normalsize{Photon flux (0.5--8.0 keV) vs. hard X-ray color (HR2) for confirmed massive-star counterparts to $Chandra$ X-ray sources. Sources not detected at energies above 3.3 keV were assigned HR2=$-$1.1 for the figure, but have HR2=$-9.00$ in Table 5. The massive stars (large dots) are systematically brighter and softer in X-rays than the field population of X-ray sources (small grey points).  Sources associated with the Arches (pentacles within circles) and Quintuplet (pentacles) clusters are marked separately for comparison. The data are accompanied by models of optically thin thermal plasma (solid lines) and non-thermal power-law emission (dotted lines) absorbed by a hydrogen column of $N_{\textrm{\scriptsize{H}}}=6.0\times10^{22}$ cm$^{-2}$. The implied X-ray luminosities of the massive stars are in the range $L_{\textrm{\scriptsize{X}}}\approx10^{32}$--$10^{34}$ erg s$^{-1}$ (0.5--8.0 keV).}}
\label{fig:hardint}
\end{figure*}

\begin{figure*}[t]
\centering
\epsscale{1}
\plottwo{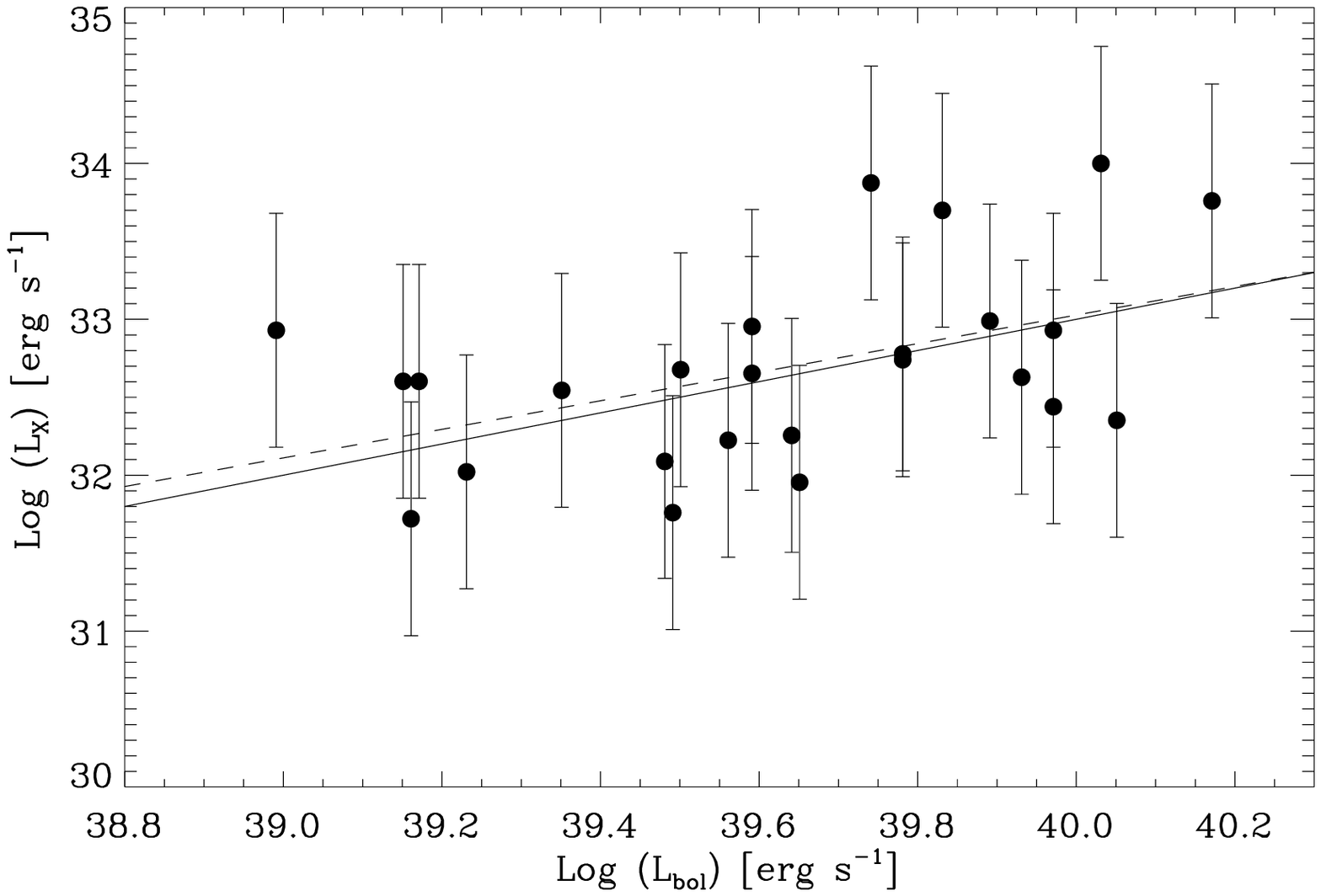}{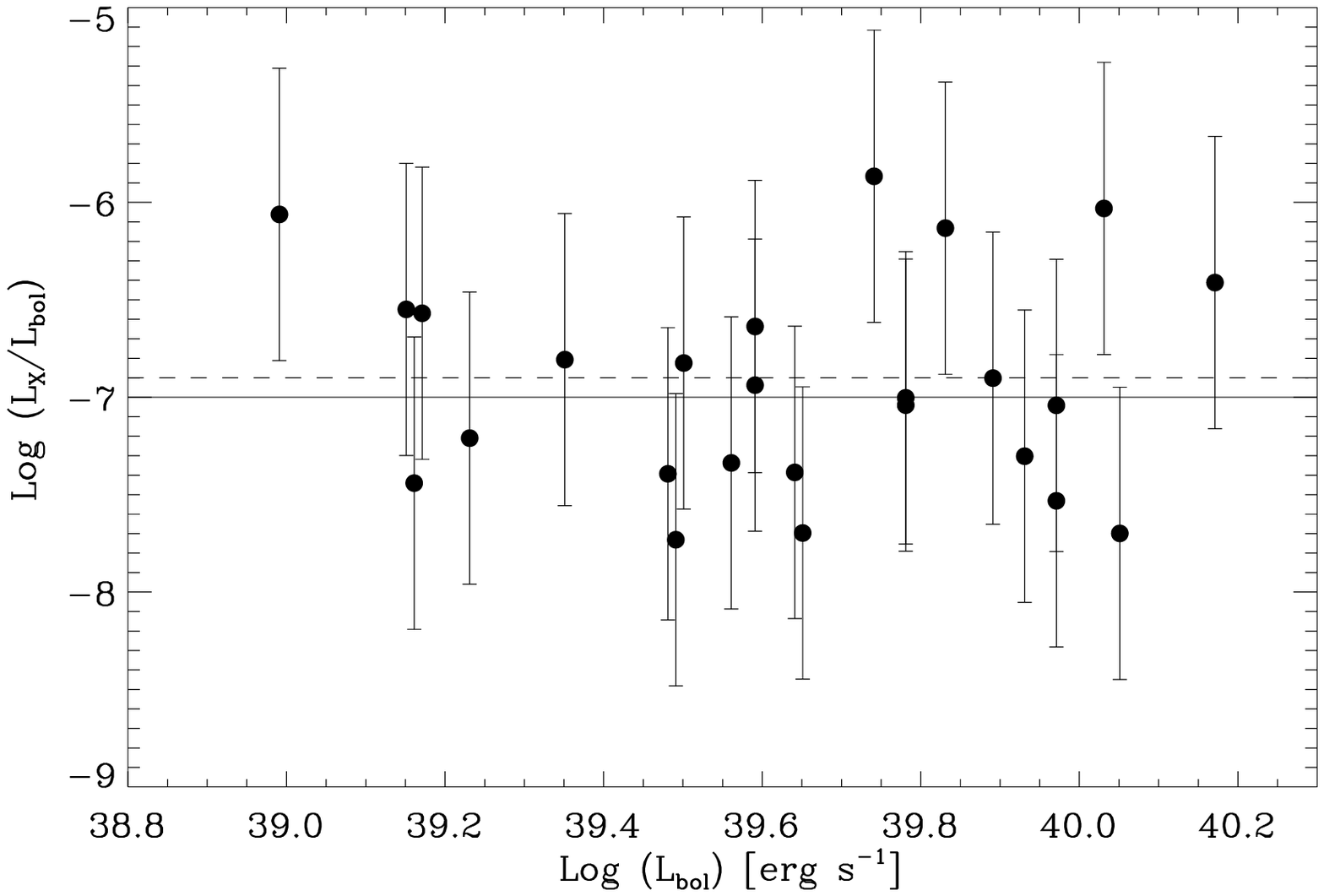}
\caption[X-ray/bolometric luminosity relations]{\linespread{1}\normalsize{Scatter plots illustrating the relationship between the X-ray luminosity (0.5--8.0 keV) and the stellar bolometric luminosity of the massive stellar X-ray sources near the Galactic center. A linear fit to the data (dashed line) yields a value of $L_{\textrm{\scriptsize{X}}}/L_{\textrm{\scriptsize{bol}}}=10^{-6.9\pm0.6}$, which is consistent with the canonical value of $10^{-7}$ (solid line). }}
\end{figure*}

\begin{figure*}[t]
\centering
\epsscale{1}
\plotone{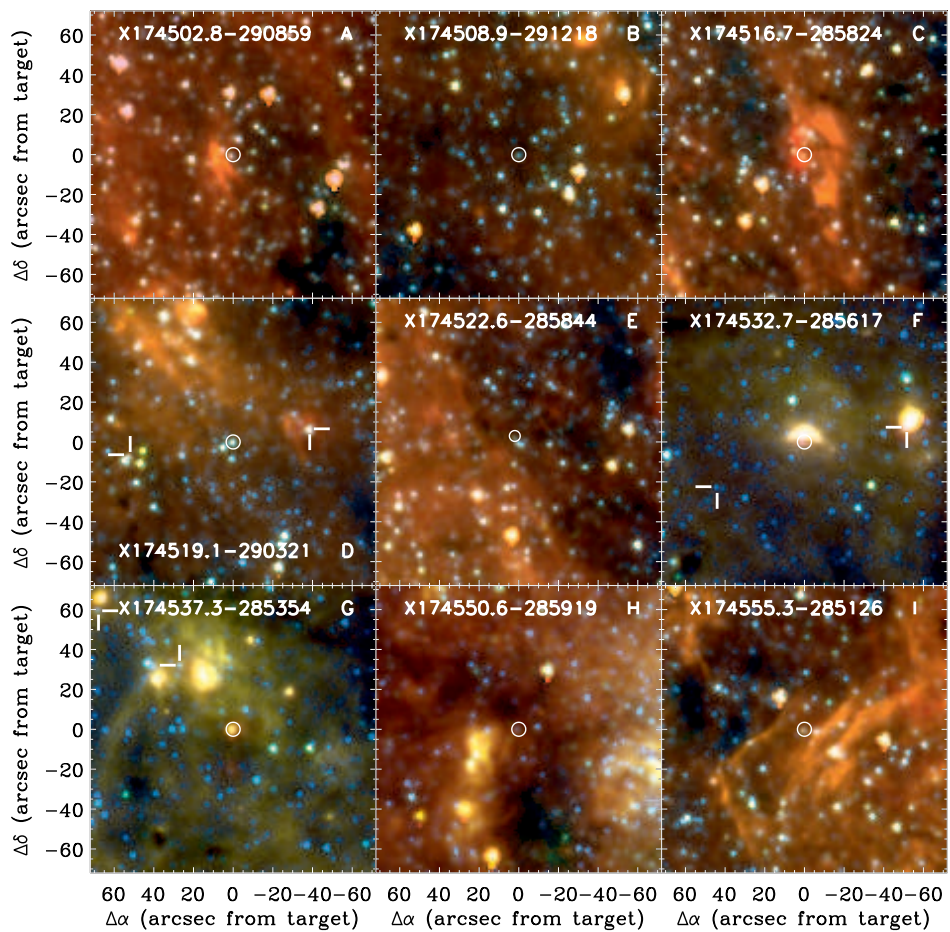}
\caption[$Spitzer$/IRAC Images of Massive Stars Environs 1]{\linespread{1}\normalsize{$Spitzer$/IRAC images of fields containing massive stellar X-ray sources. Red, green, and blue correspond to 8.0, 5.8, and 3.6 {\micron} images, respectively. Each image spans 144\arcsec~on a side, and is centered on the labeled sources. Other known massive stars, if present within these fields, are marked by perpendicular pointers. In all images, north is up and east is to the left. Image contrast was scaled by histogram equalization to help reveal faint structure, so the colors of the images are not necessarily illustrative of source SEDs.}}
\end{figure*}

\begin{figure*}[t]
\centering
\epsscale{1}
\plotone{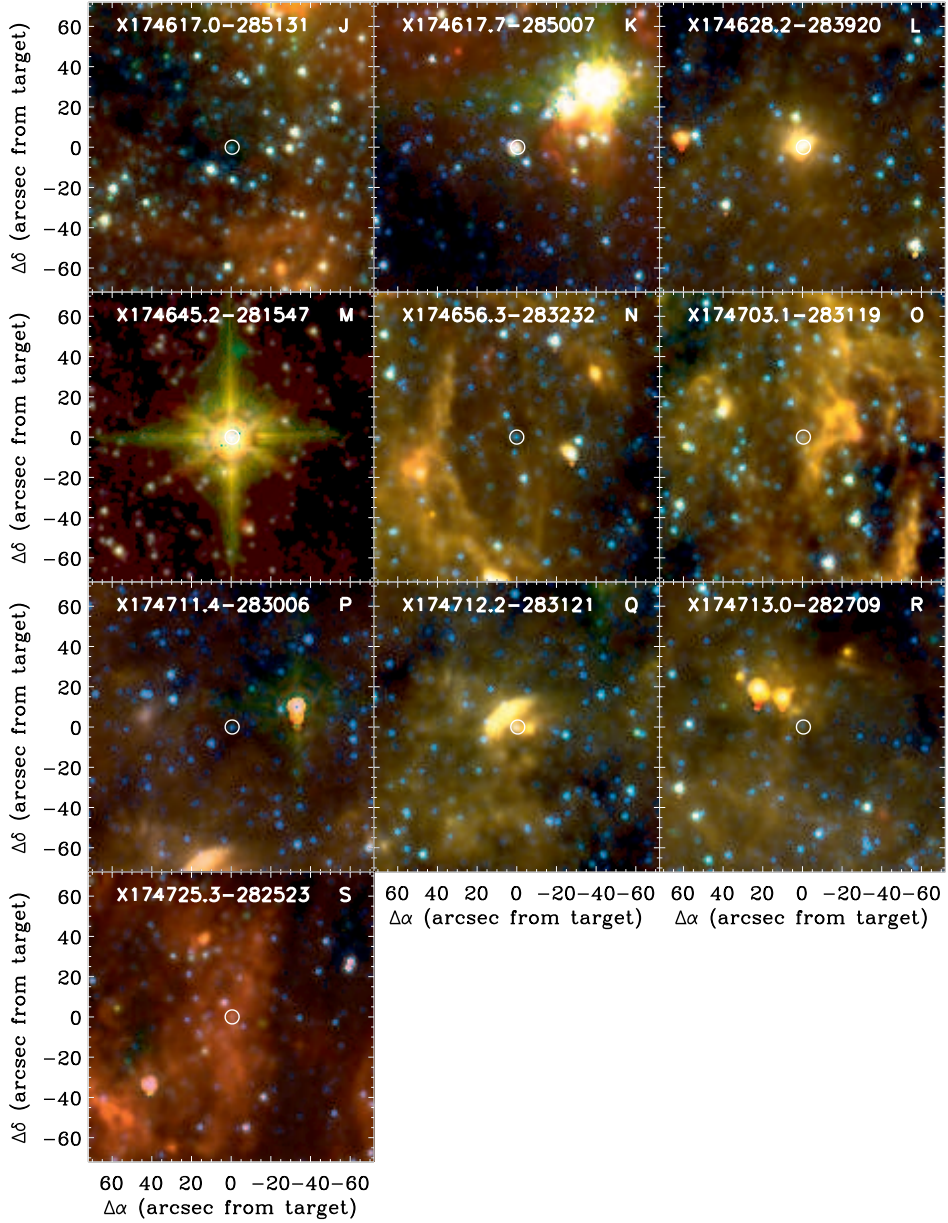}
\caption[$Spitzer$/IRAC images of Massive Star Environs 2]{\linespread{1}\normalsize{$Spitzer$/IRAC images of fields containing additional massive stellar X-ray sources.}}
\end{figure*}

\begin{landscape}
\begin{figure*}[t]
\centering
\epsscale{1}
\plotone{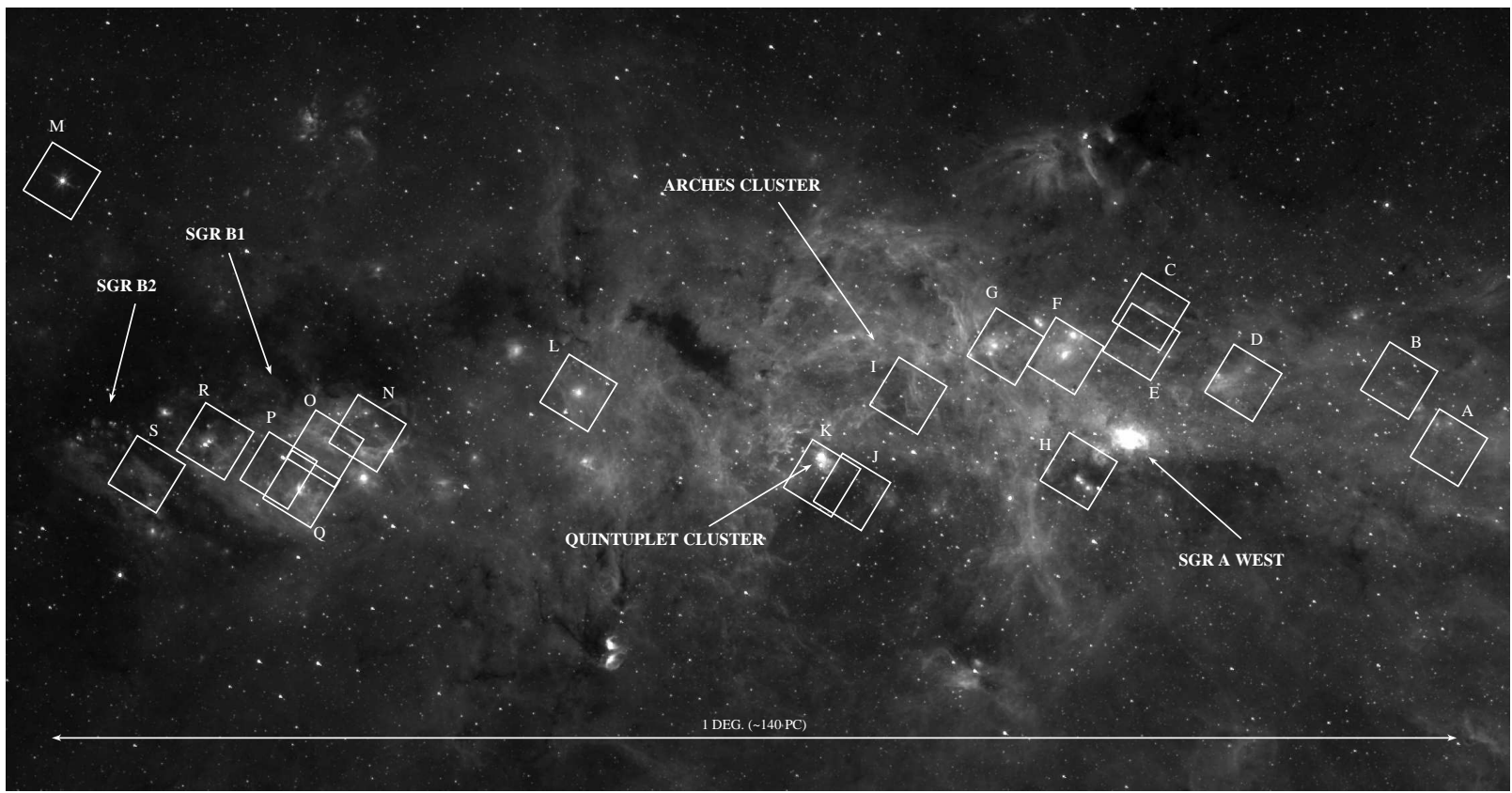}
\caption[\textit{Spitzer}/IRAC $\lambda$8 {\micron} Image of Galactic Center, New Massive Stars]{\linespread{1}\normalsize{\textit{Spitzer}/IRAC $\lambda$8 {\micron} image of the Galactic center region (Stolovy et al. 2006; Arendt et al. 2008), oriented so the Galactic plane is horizontal. The field of each massive stellar X-ray source studied in this work is marked, matching the individual, zoomed color images in Figures 13 and 14.}}
\end{figure*}
\end{landscape}

\end{document}